\documentclass[prd,twocolumn,aps,floatfix,showpacs,nofootinbib]{revtex4}
\usepackage{graphicx}
\usepackage{latexsym}
\usepackage{revsymb}
\usepackage{amsfonts}
\usepackage{amsmath}
\usepackage{amssymb}
\usepackage{bm}

\newcommand{\vev}[1]{\langle #1\rangle}
\newcommand{\ket}[1]{\left| #1 \right\rangle}
\newcommand{\bra}[1]{\left\langle #1 \right|}

\newcommand{\arcsinh}{\mathrm{arcsinh}}
\newcommand{\arccosh}{\mathrm{arccosh}}
\newcommand{\iomega}{\omega}

\begin{document}

\title{Position and Momentum Uncertainties of the Normal and Inverted Harmonic Oscillators
under the Minimal Length Uncertainty Relation}

\author{Zachary Lewis}\email{zlewis@vt.edu}
\affiliation{Department of Physics, Virginia Tech, Blacksburg, VA 24061, USA}
\author{Tatsu Takeuchi}\email{takeuchi@vt.edu}
\affiliation{Department of Physics, Virginia Tech, Blacksburg, VA 24061, USA}

\date{September 11, 2011}

\begin{abstract}
We analyze the position and momentum uncertainties of the energy eigenstates of 
the harmonic oscillator in the context of a deformed quantum mechanics, 
namely, that in which the commutator between the position and momentum operators
is given by $[\hat{x},\hat{p}]=i\hbar(1+\beta \hat{p}^2)$.
This deformed commutation relation leads to the minimal length uncertainty
relation $\Delta x\ge (\hbar/2)(1/\Delta p +\beta\Delta p)$, 
which implies that $\Delta x\sim 1/\Delta p$ at small $\Delta p$ while
$\Delta x \sim \Delta p$ at large $\Delta p$. We find that the
uncertainties of the energy eigenstates of the normal harmonic oscillator ($m>0$), derived in Ref.~\cite{Chang:2001kn},
only populate the $\Delta x\sim 1/\Delta p$ branch.
The other branch, $\Delta x\sim \Delta p$, is found to be populated by the
energy eigenstates of the `inverted' harmonic oscillator ($m<0$). The Hilbert space in the 'inverted' case admits an infinite ladder of positive energy 
eigenstates provided that 
$\Delta x_{\min}  = \hbar\sqrt{\beta} > \sqrt{2}\,[\,\hbar^2/k|m|\,]^{1/4}$.
Correspondence with the classical limit is also discussed.
\end{abstract}


\maketitle

\section{Introduction}

The consequences of deforming the canonical commutation relation between 
the position and momentum operators to
\begin{equation}
[\, \hat{x},\, \hat{p} \,] \;=\; i\hbar (1 + \beta \hat{p}^2) 
\label{CommutationRelation}
\end{equation}
have been studied in various contexts 
by many authors \cite{Chang:2001kn,Kempf:1997fz,Brau:1999uv,Chang:2001bm,Hossenfelder:2003jz,Das:2008kaa,Bagchi:2009wb,Pedram:2010hx}.
The main motivation behind this was to use such deformed
quantum mechanical systems as models which obey the
minimal length uncertainty relation (MLUR) \cite{Amati:1988tn}:
\begin{equation}
\Delta x \;\ge\; \frac{\hbar}{2}
             \left( \frac{1}{\Delta p} + \beta\,\Delta p
             \right)\;.
\label{MLUR}
\end{equation}
This relation is expected on fairly generic grounds in quantum gravity \cite{Wheeler:1957mu,MinimalLength}, 
and has been observed in perturbative string theory \cite{gross}.
The MLUR implies the existence of a minimal length
\begin{equation}
\Delta x_{\min} \;=\; \hbar\sqrt{\beta}\;,
\label{MinLength}
\end{equation}
below which the uncertainty in position, $\Delta x$, 
cannot be reduced.
For quantum gravity, $\Delta x_{\min}$ would be the Planck scale, $\ell_\mathrm{P}=\sqrt{\hbar G_N/c^3}$,
while for string theory this would be the string length scale, $\ell_s=\sqrt{\alpha'}$.
The investigation of said model systems could be expected to shed some light
on the nature of these, and other, theories which possess a minimal length scale.

\begin{figure}[t]
\includegraphics[width=8cm]{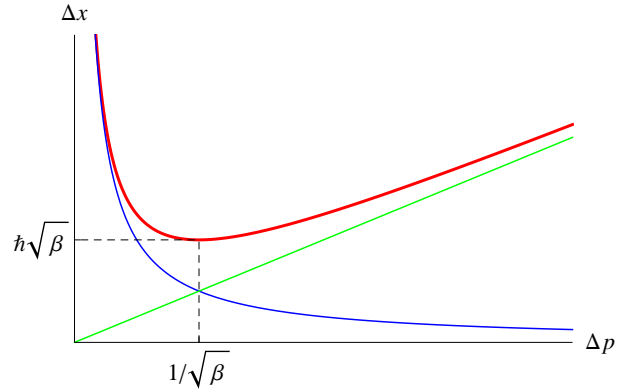}
\caption{The lower bound on $\Delta x$ as a function of $\Delta p$ when they
obey the minimal length uncertainty relation, Eq.~(\ref{MLUR}), is shown in red.
The blue line indicates $\Delta x = (\hbar/2)(1/\Delta p)$ while the green line
indicates $\Delta x = (\hbar\beta/2)\Delta p$.
}
\label{MLURfig}
\end{figure}

Note that the uncertainty in position which saturates the MLUR bound
behaves as $\Delta x \sim 1/\Delta p$ for $\Delta p < 1/\sqrt{\beta}$, while
$\Delta x \sim \Delta p$ for $1/\sqrt{\beta} < \Delta p$, as illustrated 
in Fig.~\ref{MLURfig}.
While we are familiar with the $\Delta x \sim 1/\Delta p$ behavior from
canonical quantum mechanics, the $\Delta x \sim \Delta p$ behavior is quite
novel. It behooves us to understand how it can come about, and how
it can coexist with the canonical $\Delta x \sim 1/\Delta p$ behavior
within a single quantum mechanical system.
To this end, we calculate $\Delta x$ and $\Delta p$ for the
energy eigenstates of the harmonic oscillator,
\begin{equation}
\hat{H} \;=\; \dfrac{1}{2}k\hat{x}^2 + \dfrac{1}{2m}\,\hat{p}^2\;,
\end{equation}
the wave-functions of which were derived explicitly
in Ref.~\cite{Chang:2001kn}.
We find that all of the eigenstates of this
Hamiltonian inhabit the $\Delta x\sim 1/\Delta p$ branch, and not  
the other branch, as long as both the spring constant $k$ and the mass $m$ 
remain positive.
The cross-over happens when the inverse of the mass, $1/m$, is allowed to decrease 
through zero into the negative, in which case all the energy eigenstates
will move smoothly over to the $\Delta x\sim \Delta p$ branch.
The objective of this paper is to provide a detailed account of
this result. 

In the following, we solve the Schr\"odinger equation for the
above Hamiltonian without assuming a specific sign for the mass $m$.
The spring constant $k$ is kept positive throughout.
We find that the `inverted' harmonic oscillator with $k>0$ and $m<0$
admits an infinite ladder of normalizable positive energy eigenstates 
provided that
\begin{equation}
\Delta x_{\min} \;=\; \hbar\sqrt{\beta} \;>\; \sqrt{2}\,a\;,
\label{NegativeMassCondition}
\end{equation}
where 
\begin{equation}
a \;=\; \left[\dfrac{\hbar^2}{k|m|}\right]^{1/4}
\label{adef}
\end{equation}
is the characteristic length scale of the harmonic oscillator.
The uncertainties $\Delta x$ and $\Delta p$ are calculated for
the energy eigenstates, and the above mentioned cross-over through $1/m=0$
from the $\Delta x\sim 1/\Delta p$ branch to the $\Delta x\sim\Delta p$ branch
is demonstrated.

We then take the classical limit of our deformed commutation relation and
work out the evolution of the classical harmonic oscillator for both the positive and
negative mass cases.  
It is found that for the `inverted' $m<0$ case, the time it takes for the particle
to travel from $x=-\infty$ to $x=+\infty$ is finite, demanding the compactification of
$x$-space, and also rendering the classical probability of finding the particle
near the origin finite.
This provides a classical explanation of why `bound' states are possible for 
the `inverted' harmonic oscillator in this modified mechanics.

\section{Quantum States and Uncertainties}

\subsection{Eigenvalues and Eigenstates}

\begin{figure}[t]
\includegraphics[width=8cm]{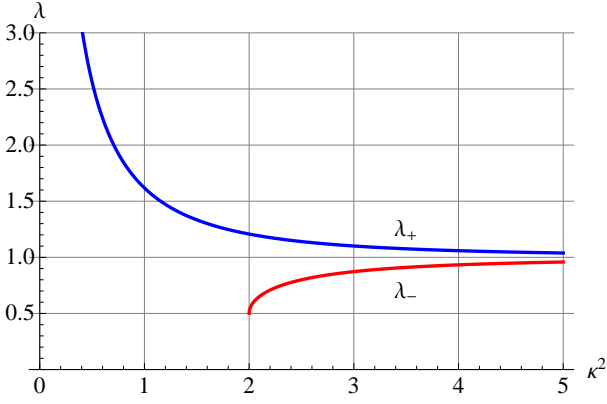}
\caption{Plots of $\lambda_+$ (blue) and $\lambda_-$ (red), Eq.~(\ref{lambdaPlusMinus}), as 
functions of $\kappa^2 = \beta\hbar\sqrt{k|m|}$.}
\label{lambdaPlot}
\end{figure}

The position and momentum operators obeying Eq.~(\ref{CommutationRelation})
can be represented in momentum space by \cite{Kempf:1995su}
\begin{eqnarray}
\hat{x} & = & i\hbar\,(1+\beta p^2)\,\frac{\partial}{\partial p}\;,  \cr
\hat{p} & = & p\;.
\label{Rep1D}
\end{eqnarray}
The inner product between two states is 
\begin{equation}
\langle \phi | \psi \rangle
\;=\; \int_{-\infty}^{\infty} \frac{dp}{(1 + \beta p^2)}\;\phi^*(p)\,\psi(p)\;.
\label{Product1D}
\end{equation}
This definition ensures the symmetricity of the operator $\hat{x}$.
The Schr\"odinger equation for the harmonic oscillator in this representation is
thus
\begin{equation}
\Biggl[ -\dfrac{\hbar^2 k}{2}\!
		  \left\{
          	 (1+\beta p^2)\frac{\partial}{\partial p}
          \right\}^2
     +\frac{1}{2m}\,p^2
\Biggr]\Psi(p)
\;=\; E\,\Psi(p)\;.
\label{Schrodinger-p}
\end{equation}
Here, we do not assume $m>0$ as usual, so the kinetic energy term can contribute with
either sign.
A change of variable from $p$ to
\begin{equation}
\rho \;\equiv\; \frac{1}{\sqrt{\beta}} \arctan(\sqrt{\beta}p) \;,
\label{rhodef}
\end{equation}
maps the region $-\infty < p < \infty$ to
\begin{equation}
-\frac{\pi}{2\sqrt{\beta}} < \rho < \frac{\pi}{2\sqrt{\beta}} \;,
\label{rhoRange}
\end{equation}
and casts 
the $\hat{x}$ and $\hat{p}$ operators into the forms
\begin{eqnarray}
\hat{x} & = & i\hbar\dfrac{\partial}{\partial\rho}\;,\cr
\hat{p} & = & \dfrac{1}{\sqrt{\beta}}\tan(\sqrt{\beta}\rho) \;,
\end{eqnarray}
with inner product given by
\begin{equation}
\langle \phi | \psi \rangle
\;=\; \int_{-\pi/2\sqrt{\beta}}^{\pi/2\sqrt{\beta}} d\rho\;\phi^*(\rho)\,\psi(\rho)\;.
\end{equation}
Note that $\hat{x}$ is the wave-number operator in $\rho$-space,
so the Fourier coefficients of the wave-function in $\rho$-space
will provide the probability amplitudes for a discretized $x$-space.
Eq.~(\ref{Schrodinger-p}) is thus transformed into:
\begin{equation}
\Biggl[
-\dfrac{\hbar^2 k}{2}\,\frac{\partial^2}{\partial\rho^2}
+\frac{1}{2m\beta}
  \tan^2 \sqrt{\beta}{\rho}
\Biggr]\Psi(\rho) \;=\; E\,\Psi(\rho)\;.
\label{Schrodinger-rho}
\end{equation}
In $\rho$-space, the potential energy term $k\hat{x}^2/2$ effectively becomes the
kinetic energy, and the kinetic energy term $\hat{p}^2/2m$ effectively becomes a
tangent-squared potential which is `inverted' when $1/m<0$.
We next introduce dimensionless parameters and a dimensionless variable by
\begin{eqnarray}
\kappa & \equiv & \bigl[\,\beta^2\hbar^2 k|m|\,\bigr]^{1/4}
\;=\; \dfrac{\Delta x_{\min}}{a}
\;,\cr
\varepsilon & \equiv & \frac{2|m|E\beta}{\kappa^2}
\;,\cr
\xi & \equiv & \frac{\sqrt{\beta}\rho}{\kappa}
\;,
\end{eqnarray}
where the length-scale $a$ was introduced in Eq.~(\ref{adef}).
The dimensionless variable $\xi$ is in the range 
\begin{equation}
-\dfrac{\pi}{2\kappa} < \xi < \dfrac{\pi}{2\kappa}\;,
\end{equation}
and the inner product is
\begin{equation}
\langle \phi | \psi \rangle 
\;=\; \dfrac{\kappa}{\sqrt{\beta}}
\int_{-\pi/2\kappa}^{\pi/2\kappa} d\xi\;\phi^*(\xi)\,\psi(\xi)\;.
\end{equation}
The dimension of the inner product has all been absorbed into the prefactor
$1/\sqrt{\beta}$.
The Schr\"odinger equation becomes
\begin{eqnarray}
& & 
\left[\; \frac{\partial^2}{\partial\xi^2}
      \mp\frac{1}{\kappa^2} 
      \tan^2\kappa\xi
      +\varepsilon \;
\right]\Psi(\xi) = 0\;,
\label{Schrodinger-xi}
\end{eqnarray}
where the minus sign in front of the tangent-squared potential is for the case $m>0$,
and the plus sign for the case $m<0$.
Let $\Psi(\xi) = c^{\lambda}\,f(s)$, where 
$s\equiv \sin\kappa\xi$, $c\equiv \cos\kappa\xi = \sqrt{1-s^2}$,
and
$\lambda$ is a constant to
be determined.  The variable $s$ is in the range
\begin{equation}
-1 < s < 1\;,
\end{equation}
with inner product given by
\begin{equation}
\langle \phi | \psi \rangle 
\;=\; \dfrac{1}{\sqrt{\beta}}
\int_{-1}^{1} \dfrac{ds}{c}\;\phi^*(s)\,\psi(s)\;.
\label{inner-s}
\end{equation}
\noindent
The equation for $f(s)$ is
\begin{eqnarray}
& &
(1-s^2)f'' - (2\lambda + 1)\,s\,f'
\cr
& &
+ \left[ \biggl\{ \frac{\varepsilon}{\kappa^2} - \lambda
         \biggr\}
       + \biggl\{ \lambda(\lambda-1)\mp\frac{1}{\kappa^4} 
         \biggr\}\frac{s^2}{c^2}
  \right] f
= 0\;.
\cr
& &
\label{Schrodinger-s1}
\end{eqnarray}
We fix $\lambda$ by requiring the coefficient of the
tangent squared term to vanish:
\begin{equation}
\lambda(\lambda-1) \mp \frac{1}{\kappa^4} = 0\;.
\end{equation}
The solutions are
\begin{equation}
\lambda 
\;=\;
\left\{
\begin{array}{ll}
\dfrac{1}{2} + \sqrt{ \dfrac{1}{4} + \dfrac{1}{\kappa^4} }
\;\equiv\;\lambda_+ \qquad & (m>0)\;, \\
\dfrac{1}{2} + \sqrt{ \dfrac{1}{4} - \dfrac{1}{\kappa^4} }
\;\equiv\;\lambda_- \qquad & (m<0)\;, \\
\end{array}
\right.
\label{lambdaPlusMinus}
\end{equation}
where we have chosen the branches for which $\lambda\ge 1/2$
to prevent the inner-product, Eq.~(\ref{inner-s}), from blowing up
at the domain boundaries.
Note that $1< \lambda_+$ while $\frac{1}{2}\le \lambda_- < 1$,
and that $\lambda_\pm\rightarrow 1$ in the limit $\kappa^2 = \beta\hbar\sqrt{k|m|}\rightarrow\infty$.
The dependence of $\lambda_\pm$ on $\kappa^2$
is shown in Fig.~\ref{lambdaPlot}.
The $\lambda_-$ branch does not extend below $\kappa^2=2$.
%
%
Setting $\lambda=\lambda_+$ for $m>0$, and $\lambda=\lambda_-$ for $m<0$
simplifies Eq.~(\ref{Schrodinger-s1}) to
\begin{equation}
(1-s^2)\,f'' - (\,2\lambda + 1\,)\,s\,f'
+ \left( \frac{\varepsilon}{\kappa^2} - \lambda \right) f
= 0\;,
\label{Schrodinger-s2}
\end{equation}
the sign of $m$ being encoded in the value of $\lambda$.
Since $f(s)$ should be non-singular at $s=\pm 1$,
we demand a polynomial solution to Eq.~(\ref{Schrodinger-s2}).  
This requirement imposes the following condition on the coefficient of $f$:
\begin{equation}
\frac{\varepsilon}{\kappa^2} - \lambda
= n \,(\, n + 2\lambda \,) \;,
\label{EigenPM1}
\end{equation}
where $n$ is a non--negative integer \cite{SpecialFunctions}.
Eq.~(\ref{Schrodinger-s2}) becomes
\begin{equation}
(1-s^2)f'' - (\, 2\lambda + 1 \,)\,s\,f'
+ n \,(\, n + 2\lambda \,)\, f = 0\;,
\label{Schrodinger-s3}
\end{equation}
the solution of which is given by the Gegenbauer polynomial:
\begin{equation}
f(s) \;=\; C_n^{\lambda}(s)\;.
\end{equation}
The Gegenbauer polynomials satisfy the following orthogonality relation:
\begin{equation}
\int_{-1}^{1} c^{2\lambda-1}\, C_n^\lambda(s)\,C_m^\lambda(s)\,ds
\;=\; \dfrac{2\pi\,\Gamma(n+2\lambda)}{[\,2^\lambda\Gamma(\lambda)\,]^2\,n!\,(n+\lambda)\,}
\,\delta_{nm}\;.
\label{GegenbauerOrthogonalityRelation}
\end{equation}
The energy eigenvalues follow from the condition Eq.~(\ref{EigenPM1}).
Replacing $\lambda$ with $\lambda_\pm$, we find
\begin{eqnarray}
\varepsilon_n^{(\pm)}
& = & \kappa^2 \left[\, n^2 + (2n+1)\lambda_\pm \,\right] \phantom{\bigg|} \cr
& = & \dfrac{n^2+(2n+1)\lambda_\pm}{\sqrt{\lambda_\pm\bigl|\lambda_\pm-1\bigr|}} \cr
& = & \kappa^2 \left( n^2 + n + \frac{1}{2} \right)
    + \left( 2n+1 \right)
      \sqrt{ \frac{\kappa^4}{4} \pm 1 } 
\;,
\cr
& &
\label{epsilonPMn}
\end{eqnarray}
or in the original dimensionful units,
\begin{eqnarray}
E_n^{(\pm)} 
& = & \dfrac{1}{2\beta|m|}\;\dfrac{n^2+(2n+1)\lambda_\pm}{\lambda_\pm\bigl|\lambda_\pm-1\bigr|} \cr
& = & \hbar\omega
\biggl[ \left( n + \frac{1}{2} \right)
	    \sqrt{ \dfrac{\beta^2 m^2\hbar^2\omega^2}{4}\pm 1 }
\cr
& &
\qquad\qquad
     + \left( n^2 + n + \frac{1}{2} \right) \dfrac{\beta|m|\hbar\omega}{2}
\biggr]
\cr
& = & \dfrac{k}{2}
\biggl[ \left( n + \frac{1}{2} \right)
        \sqrt{ (\Delta x_{\min})^4 \pm 4 a^4 }
\cr
& &
\qquad\qquad
     + \left( n^2 + n + \frac{1}{2} \right) (\Delta x_{\min})^2
\biggr]
\;,
\cr & &
\label{EnPM}
\end{eqnarray}
where $\omega=\sqrt{k/|m|}$.
For the $m>0$ case, we can take the limit 
$\Delta x_{\min}=\hbar\sqrt{\beta}\rightarrow 0$, and we recover
\begin{equation}
\lim_{\Delta x_{\min}\rightarrow 0}E_n^{(+)}
\;=\; ka^2\left(n+\dfrac{1}{2}\right)
\;=\; \hbar\omega\left(n+\dfrac{1}{2}\right)\;.
\label{betaZeroLimit}
\end{equation}
For the $m<0$ case, it is clear that we must have
$\Delta x_{\min} \ge \sqrt{2}a$ for the square-root in Eq.~(\ref{EnPM}) 
to remain real.
This is the condition we cited in Eq.~(\ref{NegativeMassCondition}).
Therefore, the limit $\Delta x_{\min}=\hbar\sqrt{\beta}\rightarrow 0$ cannot be taken 
in this case for non-zero $a$.
The two cases converge when $|m|\rightarrow\infty$, at which 
$a=0$, and we find that the energy levels in that limit are
\begin{equation}
\lim_{|m|\rightarrow\infty}E_n^{(\pm)}
\;=\; \dfrac{k}{2}(\Delta x_{\min})^2 (n+1)^2\;.
\label{mInfiniteLimit}
\end{equation}
Thus, the $1/m>0$ and $1/m<0$ cases connect smoothly at $1/m=0$.

\begin{figure}[t]
\includegraphics[width=8cm]{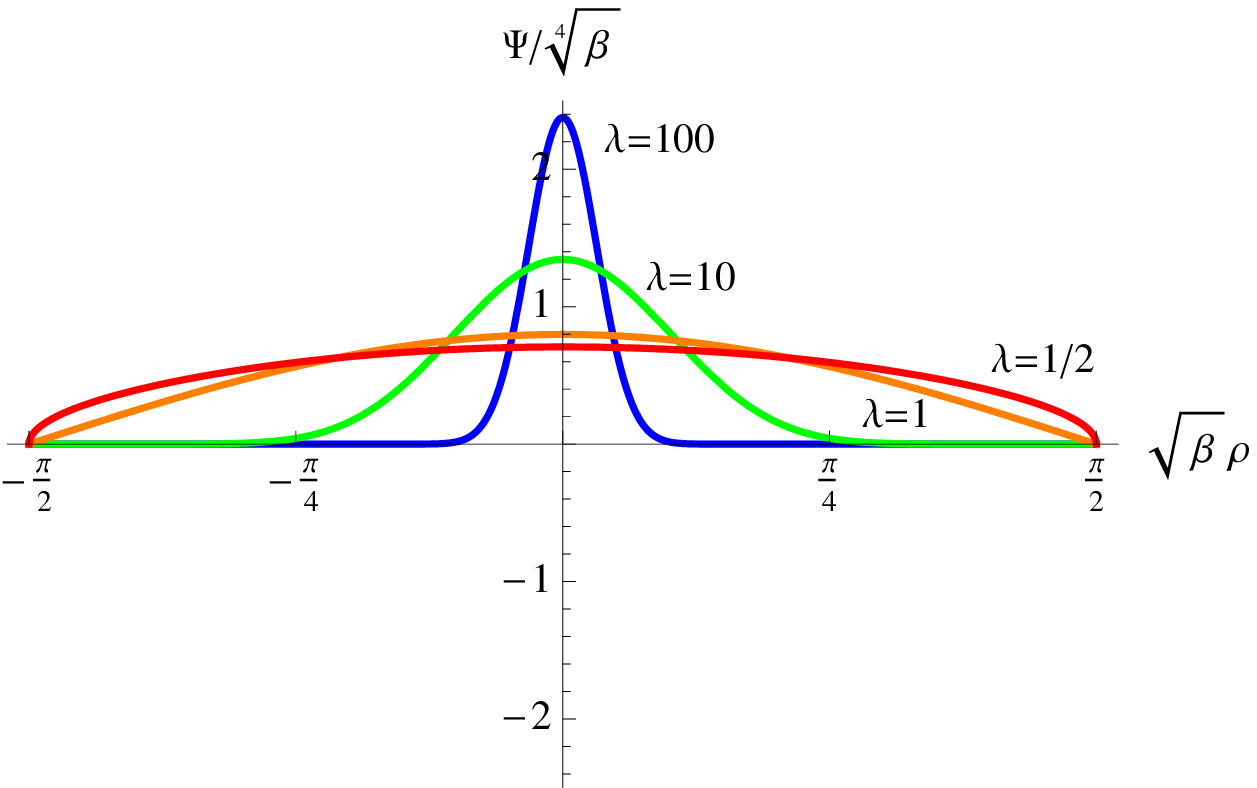}
\includegraphics[width=8cm]{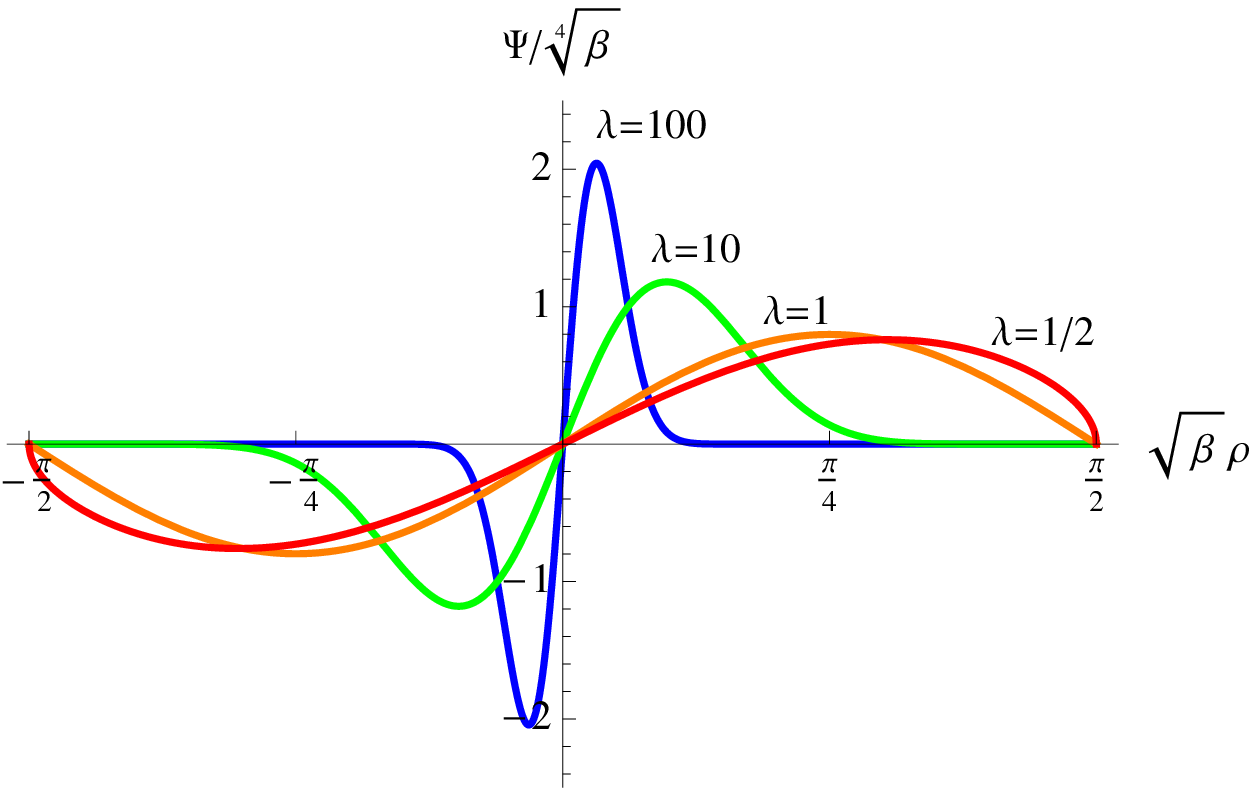}
\includegraphics[width=8cm]{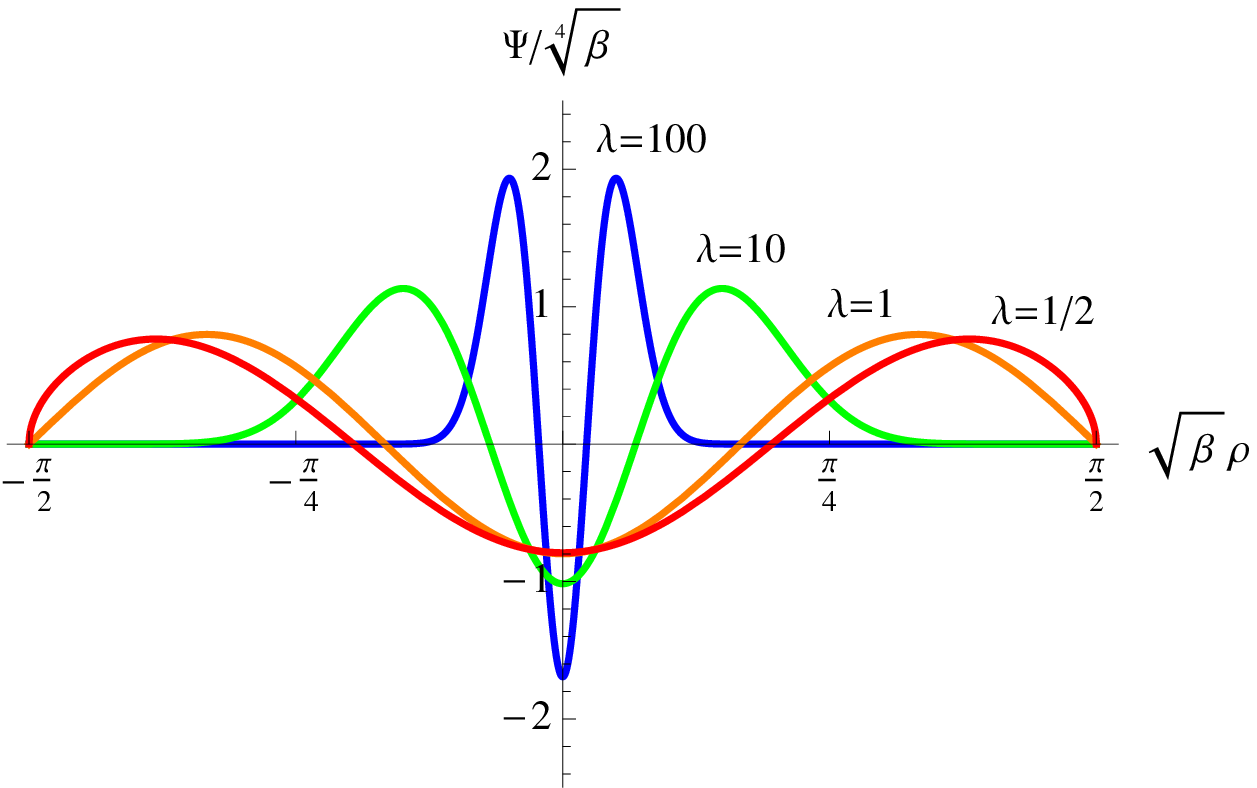}
\caption{$\lambda$-dependence of the wave-functions of the first few energy eigenstates.
The $\lambda>1$ values correspond to $m>0$, while the
$\frac{1}{2}\le\lambda<1$ values correspond to $m<0$.  
The $\lambda=1$ case corresponds to the limit $|m|\rightarrow\infty$.
}
\label{WaveFunctions}
\end{figure}

The normalized energy eigenfunctions are thus given by:
\begin{equation}
\Psi_n^{(\lambda)}(p) \;=\; N_n^{(\lambda)} \;c^{\lambda}\,C_n^{\lambda}(s)
\label{NormalizedEigenfunctions}
\;,
\end{equation}
where
\begin{eqnarray}
N_n^{(\lambda)} & = &
\sqrt[4]{\beta}\left[
2^{\lambda}\Gamma(\lambda)
\sqrt{ \dfrac{ n!\,(n+\lambda) }{ 2\pi\,\Gamma(n+2\lambda) } }
\right] \;,\cr
c & = & \cos\sqrt{\beta}\rho \;=\; \dfrac{ 1 }{ \sqrt{1+\beta p^2} }\;,\cr
s & = & \sin\sqrt{\beta}\rho \;=\; \dfrac{ \sqrt{\beta} p } 
                                              { \sqrt{1+\beta p^2} }\;.
\end{eqnarray}
The wave-functions for the first few energy eigenstates 
for several representative values of $\lambda$ are shown in Figs.~\ref{WaveFunctions}.

\subsection{Expectation Values and Uncertainties}

\begin{figure}[t]
\includegraphics[width=8cm]{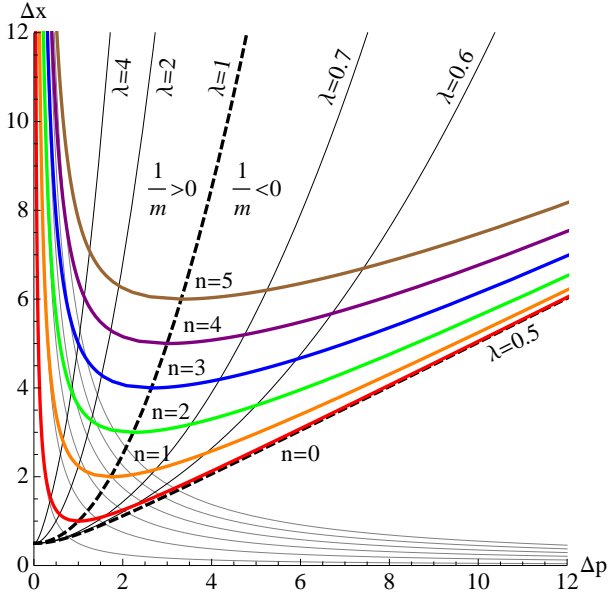}
\caption{$\Delta p$ versus $\Delta x$ for the lowest six energy eigenstates of the harmonic oscillator.
$\Delta x$ is in units of $\Delta x_\mathrm{min}=\hbar\sqrt{\beta}$, while $\Delta p$ is in units of
$1/\sqrt{\beta}=\hbar/\Delta x_\mathrm{min}$. 
The location of the state along the curves shown is determined by the value of $\lambda$
defined in Eq.~(\ref{lambdaPlusMinus}).  The $\lambda=1$ points correspond to the case
$1/m=0$.  As $1/m$ is increased to the positive side, the value of $\lambda$ will increase away from
one and the state will move toward the left along the $\Delta x \sim 1/\Delta p$ branch.
If $1/m$ is decreased into the negative, the value of $\lambda$ will decrease toward $1/2$,
and the state will move toward the right along the $\Delta x\sim\Delta p$ branch.
The $n=0$ curve (shown in red) corresponds to the minimal length uncertainty bound, Eq.~(\ref{MLUR}).
}
\label{HO-DpDx}
\end{figure}

Using the wave-functions derived above, and the formula provided in the appendix, 
the expectation values of $\hat{x}$, $\hat{p}$, $\hat{x}^2$, and $\hat{p}^2$ 
for the energy eigenstates are found to be
\begin{eqnarray}
\bra{n,\lambda}\hat{x}\ket{n,\lambda} & = & 0\;,\cr
\bra{n,\lambda}\hat{p}\ket{n,\lambda} & = & 0\;,\cr
\bra{n,\lambda}\hat{x}^2\ket{n,\lambda} & = & (\hbar^2\beta)\dfrac{(\lambda+n)\bigl[(2\lambda-1)n+\lambda\,\bigr]}{(2\lambda-1)}\;,\cr
\bra{n,\lambda}\hat{p}^2\ket{n,\lambda} & = & \dfrac{1}{\beta}\left(\dfrac{2n+1}{2\lambda-1}\right)\;,
\end{eqnarray}
giving the uncertainties in $\vev{\hat{x}}$ and $\vev{\hat{p}}$ as
\begin{eqnarray}
\Delta x_n & = & \Delta x_\mathrm{min}\sqrt{\dfrac{(\lambda+n)
\bigl[(2\lambda-1)n+\lambda\,\bigr]}{(2\lambda-1)}}\;,\cr
\Delta p_n & = & \dfrac{1}{\sqrt{\beta}}\sqrt{\dfrac{2n+1}{2\lambda-1}}\;.
\label{DxnDpn}
\end{eqnarray}
For fixed $n$ and fixed $\beta$, $\Delta x_n$ is a monotonically
decreasing function of $\lambda$ in the range $\frac{1}{2}<\lambda<1$,
and a monotonically increasing one in the range $1<\lambda$.
$\Delta p_n$ on the other hand is a monotonically decreasing function of $\lambda$ throughout.
Eliminating $\lambda$ from the above expressions, we find
\begin{eqnarray}
\dfrac{\Delta x_n}{\Delta x_\mathrm{min}}
& = & \dfrac{1}{2\sqrt{\beta}\Delta p_n}\sqrt{
\Bigl[ 1 + \beta\Delta p_n^2 \Bigr]
\Bigl[ (2n+1)^2+ \beta\Delta p_n^2 \Bigr]
} \cr
& \ge & \dfrac{1}{2}
\left( \dfrac{1}{\sqrt{\beta}\Delta p_n} + \sqrt{\beta}\Delta p_n \right)
\;, 
\label{DpDxCurve}
\end{eqnarray}
where the equality in the second line is saturated for the $n=0$ case only.
The first line gives the curve on the $\Delta p$-$\Delta x$ plane that
the point $(\Delta p_n,\Delta x_n)$ follows as $\lambda$ is varied.
Differentiating with respect to $\Delta p_n$ we find
\begin{eqnarray}
\lefteqn{
\dfrac{d}{d(\Delta p_n)}\left[\dfrac{\Delta x_n}{\Delta x_\mathrm{min}}\right]
} \cr
& = & \dfrac{ \beta^2\Delta p_n^4 - (2n+1)^2 }
{2\sqrt{\beta}\Delta p_n^2\sqrt{ 
\Bigl[ 1 + \beta\Delta p_n^2 \Bigr]
\Bigl[ (2n+1)^2+ \beta\Delta p_n^2 \Bigr]
}}
\;,
\end{eqnarray}
indicating that the curve is flat at the $\lambda=1$ point where
$\Delta p_n = \sqrt{(2n+1)/\beta}$
and $\Delta x_n$ reaches its minimum of $\Delta x_\mathrm{min}(n+1)$.
Therefore, the $\lambda=1$ point is the turn-around point where the
uncertainties switch from the $\Delta x\sim 1/\Delta p$ behavior to
the $\Delta x\sim \Delta p$ behavior. To go from one branch to another
one must flip the sign of the mass $m$.

Eliminating $n$ from Eq.~(\ref{DxnDpn}), we find
\begin{equation}
\dfrac{\Delta x_n}{\Delta x_\mathrm{min}}
\;=\; \dfrac{1}{2}\sqrt{
\left( 1 + \beta \Delta p_n^2 \right)
\Bigl[ 1 + (2\lambda-1)^2\beta\Delta p_n^2 \Bigr]
}\;.
\end{equation}
This gives the curve on which the points $(\Delta p_n,\Delta x_n)$ fall on 
for constant $\lambda$.  In particular, for $\lambda=1$ this reduces to
\begin{equation}
\dfrac{\Delta x_n}{\Delta x_\mathrm{min}}
\;=\; \dfrac{1+\beta\Delta p_n^2}{2}\;,
\label{MaginotLine}
\end{equation}
and gives the $1/m=0$ boundary between the $1/m>0$ and $1/m<0$ regions
in $\Delta x$-$\Delta p$ space.
These properties of the uncertainties have been
plotted in Fig.~\ref{HO-DpDx}.

\subsection{Limiting Cases}

As shown in Fig.~\ref{HO-DpDx}, the value of $\lambda$ determines
where the uncertainties $(\Delta p_n,\Delta x_n)$
are along their trajectories given by Eq.~(\ref{DpDxCurve}),
with $\lambda>1$ keeping the uncertainties on the
$\Delta x\sim 1/\Delta p$ branch of the trajectory,
while $\frac{1}{2}<\lambda<1$ keeping them on the
$\Delta x\sim \Delta p$ branch.
Let us consider a few limiting values of $\lambda$ to see the
behavior of the solutions there.

\subsubsection{$\lambda\rightarrow\infty$}

The $\beta\rightarrow 0$ limit only exist for the $m>0$ case
where $\lambda=\lambda_+$.
As $\beta\rightarrow 0$, the parameter 
$\lambda_+$ diverges to infinity as
\begin{equation}
\lambda \;=\; \lambda_+ \;\sim\; \dfrac{1}{m\hbar\omega\beta}
\;\xrightarrow{\beta\rightarrow 0}\;\infty\;,
\end{equation}
where $\omega=\sqrt{k/m}$.
In that limit, the Gegenbauer polynomials become Hermite polynomials:
\begin{equation}
\lim_{\lambda\rightarrow\infty} n!\,\lambda^{-n/2}\,C_n^\lambda(x/\sqrt{\lambda}) \;=\; H_n(x)\;.
\label{gegen1}
\end{equation}
Noting that as $\beta\rightarrow 0$, we have
\begin{equation}
s
\;\sim\;\sqrt{\beta}p
\;\sim\; \dfrac{p}{\sqrt{\lambda m\hbar\omega}}\;,
\end{equation}
we can conclude that
\begin{equation}
\lim_{\lambda\rightarrow\infty} n!\,\lambda^{-n/2}\,C_n^\lambda(s)
\;=\; H_n\!\left(\dfrac{p}{\sqrt{m\hbar\omega}}\right)\;.
\end{equation}
%
Similarly,
\begin{equation}
\lim_{\lambda\rightarrow\infty} c^{\lambda}
\;=\;
\lim_{\lambda\rightarrow\infty}\left(1-\dfrac{p^2}{\lambda m\hbar\omega}\right)^{\lambda/2}
\;=\;
\exp\left(-\dfrac{p^2}{2m\hbar\omega}\right)\;.
\end{equation}
Using Stirling's formula
\begin{equation}
\Gamma(z) \;\sim\;
\sqrt{2\pi}\,e^{-z}z^{z-(1/2)} + \cdots
\end{equation}
the normalization constant can be shown to converge to
\begin{eqnarray}
\lim_{\lambda\rightarrow\infty}
N_n^{(\lambda)}\,\dfrac{\lambda^{n/2}}{n!}
& = & \dfrac{1}{\sqrt{2^n n!}} \dfrac{1}{\sqrt[4]{m\hbar\omega\pi}}
\;.
\end{eqnarray}
Therefore,
\begin{eqnarray}
\lefteqn{
\lim_{\lambda\rightarrow\infty}\Psi^{(\lambda)}_n(p)
} 
\cr & = &
\dfrac{1}{\sqrt{2^n n!}}\dfrac{1}{\sqrt[4]{m\hbar\omega\pi}}
\exp\left(-\dfrac{p^2}{2m\hbar\omega}\right)
H_n\!\left(\dfrac{p}{\sqrt{m\hbar\omega}}\right)\;,
\cr & &
\end{eqnarray}
which are just the usual harmonic oscillator wave-functions
in momentum space.
The energy eigenvalues reduce to the usual
ones given in Eq.~(\ref{betaZeroLimit}).
%
%
The uncertainties reduce to the usual ones as well
\begin{eqnarray}
\Delta x_n & \xrightarrow{\lambda\rightarrow\infty} & 
\sqrt{\dfrac{\hbar}{m\omega}\left(n+\dfrac{1}{2}\right)}\;,\cr
\Delta p_n & \xrightarrow{\lambda\rightarrow\infty} &
\sqrt{\hbar m\omega\left(n+\dfrac{1}{2}\right)}\;,
\end{eqnarray}
which satisfy
\begin{equation}
\Delta x_n \Delta p_n \;\xrightarrow{\lambda\rightarrow\infty}\;
\hbar\left(n+\dfrac{1}{2}\right)
\;\ge\; \dfrac{\hbar}{2}\;.
\end{equation}

\subsubsection{$\lambda\rightarrow 1$}

The limit $\lambda=1$ is reached when $\beta$ and $k$ are kept constant 
while $|m|$ is taken to infinity.
When $\lambda=1$, the Gegenbauer polynomials become
the Chebycheff polynomials of the second kind:
\begin{equation}
C_n^1(s) \;=\; U_n(s)\;,
\label{gegen2}
\end{equation}
where
\begin{equation}
U_n(\cos\theta)\;=\;\dfrac{\sin(n+1)\theta}{\sin\theta}\;,
\label{Chebycheff1}
\end{equation}
while the normalization constant reduces to
\begin{equation}
N_n^{(\lambda)} \;\xrightarrow{\lambda\rightarrow 1}\; 
\sqrt[4]{\beta} \sqrt{\dfrac{2}{\pi}}\;.
\end{equation}
Therefore,
\begin{equation}
\dfrac{\Psi^{(1)}_n(p)}{\sqrt[4]{\beta}}
\;=\; \sqrt{\dfrac{2}{\pi}}\,c\,U_n(s)\;.
\end{equation}
The orthonormality relation for the Chebycheff polynomials is 
\begin{equation}
\int_{-1}^{1}\sqrt{1-x^2}\,U_n(x)\,U_m(x)\,dx\;=\;\dfrac{\pi}{2}\,\delta_{nm}\;,
\end{equation}
and we can see that the correct normalization constant is obtained.
Since the argument in our case is $s=\sin\sqrt{\beta}\rho$, it is more convenient to
express the Chebycheff polynomials as
\begin{eqnarray}
U_{2n}(\sin\theta) & = & (-1)^{n}\dfrac{\cos[(2n+1)\theta]}{\cos\theta}\;,\cr
\cr
U_{2n+1}(\sin\theta) & = & (-1)^n\dfrac{\sin[(2n+2)\theta]}{\cos\theta}\;,
\label{Chebycheff2}
\end{eqnarray}
for $n=0,1,2,\cdots$. 
This will allow us to write
\begin{equation}
\begin{array}{ll}
\dfrac{\Psi_{2n}^{(1)}(p)}{\sqrt[4]{\beta}}
& = \; (-1)^n \sqrt{\dfrac{2}{\pi}}\cos\Bigl[(2n+1)\sqrt{\beta}\rho\Bigr] \;,\\
\dfrac{\Psi_{2n+1}^{(1)}(p)}{\sqrt[4]{\beta}}
& = \; (-1)^n \sqrt{\dfrac{2}{\pi}}\sin\Bigl[(2n+2)\sqrt{\beta}\rho\Bigr] \;.
\end{array}
\end{equation}
%
%
%
%
%
%
The energy eigenvalues in this limit were given in Eq.~(\ref{mInfiniteLimit}).
Here, our procedure of keeping the spring constant $k$ fixed while taking $|m|$ to
infinity maintains the finiteness of $E_n$, while taking the kinetic energy 
contribution to $E_n$ to zero.
From the $n$-dependence of the energies, we can see
that, in this limit, the problem reduces to that of an infinite square well potential,
of width $\pi/\sqrt{\beta}$, in $\rho$-space.
Indeed, the effective potential in $\rho$-space was
\begin{equation}
\dfrac{1}{2m\beta}\tan^2\bigl(\sqrt{\beta}\rho\bigr)
\;\xrightarrow{m\rightarrow\infty}\;
\left\{\begin{array}{ll}
0            & \mbox{for $\rho\neq\pm\dfrac{\pi}{2\sqrt{\beta}}$} \;, \\
\infty\qquad & \mbox{at $\rho =  \pm\dfrac{\pi}{2\sqrt{\beta}}$} \;.
\end{array}
\right.
\end{equation}
This can also be seen from the form of the energy eigenfunctions, which have reduced to 
simple sines and cosines.
We will see in the next section that the classical solution also behaves 
as that of a particle in an infinite square well potential in $\rho$-space
in the same limit.

The uncertainties become
\begin{eqnarray}
\Delta x_n & \xrightarrow{\lambda\rightarrow 1} & 
\Delta x_\mathrm{min} \bigl(n+1\bigr)\;,\cr
\Delta p_n & \xrightarrow{\lambda\rightarrow 1} &
\sqrt{\dfrac{(2n+1)}{\beta}}\;,
\end{eqnarray}
as was shown in Fig.~\ref{HO-DpDx}.
Note that
\begin{eqnarray}
\dfrac{\hbar}{2}\left[\dfrac{1}{\Delta p_n} + \beta\Delta p_n\right]
& = & \Delta x_\mathrm{min}\;\dfrac{n+1}{\sqrt{2n+1}} \cr
& \le & \Delta x_\mathrm{min}\bigl(n+1\bigr) 
\;=\; \Delta x_n\;,
\cr & &
\end{eqnarray}
the bound being saturated only for the ground state $n=0$.

\subsubsection{$\lambda\rightarrow \frac{1}{2}$}

The $\lambda\rightarrow\frac{1}{2}$ limit is reached as $\Delta x_{\min}\rightarrow \sqrt{2}a$
when $m<0$. In this limit, the Gegenbauer polynomials become the Legendre polynomials,
$C_n^{1/2}(s) \;=\; P_n(s)$, while the normalization constant reduces to
\begin{equation}
N_n^{(\lambda)}\;\xrightarrow{\lambda\rightarrow\frac{1}{2}}\;
\sqrt[4]{\beta}\sqrt{\dfrac{2n+1}{2}}\;.
\end{equation}
The wavefunctions are
\begin{equation}
\dfrac{\Psi^{(1/2)}_n(p)}{\sqrt[4]{\beta}} \;=\; 
\sqrt{\dfrac{2n+1}{2}}\sqrt{c}\,P_n(s)\;. 
\end{equation}
Note that the orthonormality relation for the Legendre polynomials is
\begin{equation}
\int_{-1}^{1} P_n(x)\,P_m(x)\,dx \;=\; \dfrac{2}{2n+1}\,\delta_{nm}\;,
\end{equation}
so these wave-functions are properly normalized.
The integrals for
$\vev{\hat{x}^2}$ and $\vev{\hat{p}^2}$ diverge in this limit, 
so both $\Delta x_n$ and $\Delta p_n$ are divergent for all $n$.
However, the energy, which is the difference between
$k\vev{\hat{x}^2}/2$ and $\vev{\hat{p}^2}/2|m|$, stays finite:
\begin{equation}
E_n^{-}
\;=\; \dfrac{k}{2}(\Delta x_{\min})^2
\left(n^2+n+\dfrac{1}{2}\right)
\;.
\end{equation}
%

\section{Classical States and Uncertainties}

As we have seen above, for values of $\beta$ which maintain
the inequality $\Delta x_{\min}=\hbar\sqrt{\beta}>\sqrt{2}a$, the harmonic
oscillator Hamiltonian admits an infinite ladder of
positive energy eigenstates even when $m<0$.
Furthermore, these are states with finite $\Delta x$ and $\Delta p$,
implying that the particle is `bound' close to the phase space origin,
just as in the $m>0$ case.
But how can a particle be `bound' for an `inverted' harmonic oscillator?
To gain insight into this question,
we solve the corresponding classical equation of motion.

\subsection{The Classical Equations of Motion}

We assume that the classical limit of our commutation relation, Eq.~(\ref{CommutationRelation}),
is obtained by the usual correspondence between commutators and Poisson brackets:
\begin{equation}
\dfrac{1}{i\hbar}[\,\hat{A},\,\hat{B}\,]\;\rightarrow\;\{A,\,B\,\}\;.
\end{equation}
Therefore, we assume
\begin{eqnarray}
\{\,x,\,x\,\} & = & 0\;,\cr
\{\,p,\,p\,\} & = & 0\;,\cr
\{\,x,\,p\,\} & = & (1+\beta p^2)\;.
\label{ModifiedPoisson}
\end{eqnarray}
Then, the equations of motion for the harmonic oscillator with 
Hamiltonian given by 
\begin{equation}
H \;=\; \frac{1}{2}k x^2 + \frac{1}{2m}p^2 \;,
\label{HOHamiltonian}
\end{equation}
are
\begin{eqnarray}
\dot{x} & = & \{\,x,\,H\,\} \;=\; \dfrac{1}{m}(1+\beta p^2)\,p\;,\cr
\dot{p} & = & \{\,p,\,H\,\} \;=\; -k(1+\beta p^2)\,x\;.
\end{eqnarray}
We allow $m$ to take on either sign: if $m>0$, then $\dot{x}$ and $p$ will have the same
sign; if $m<0$ they will have opposite sign.
Note that, even though the equations of motion of $x$ and $p$ have changed, the total
energy will still be conserved.  Consequently, the time-evolution of $x$ and $p$
in phase space will be along the trajectory given by $H=\mathrm{constant}$.
For the $m>0$ case this will be an ellipse, while for the $m<0$ case this will
be a hyperbola.

To solve these equations, we change the variable $p$ to $\rho$, which was introduced in 
Eq.~(\ref{rhodef}) for the quantum case.
Then, the equations become
\begin{eqnarray}
\dot{x} & = & \dfrac{1}{m\sqrt{\beta}}\left[\dfrac{\tan(\sqrt{\beta}\rho)}{\cos^2(\sqrt{\beta}\rho)}\right]
\;=\; \dfrac{1}{2m\beta}\dfrac{d}{d\rho}\left[\tan^2(\sqrt{\beta}\rho)\right]\;, \cr
\dot{\rho} & = & -k x\;.\phantom{\dfrac{X}{X}}
\end{eqnarray}
Therefore,
\begin{equation}
\ddot{\rho} 
\;=\; -k \dot{x} 
\;=\; -\dfrac{k}{2m\beta}\,\dfrac{d}{d\rho}\left[\tan^2(\sqrt{\beta}\rho)\right]\;,
\end{equation}
%
which integrates to
\begin{equation}
\dot{\rho}^2 \;=\; -\dfrac{k}{m\beta}
\biggl[\,\tan^2\bigl(\sqrt{\beta}\rho\bigr) - C \,\biggr]\;,
\label{rhodotsquared}
\end{equation}
where $C$ is the integration constant.
Since we must have $\dot{\rho}^2>0$, the range of
allowed values of $C$ will depend on whether $m>0$ or $m<0$.
We will consider the two cases separately.


\begin{figure}[t]
\begin{flushright}
\includegraphics[width=7.84cm]{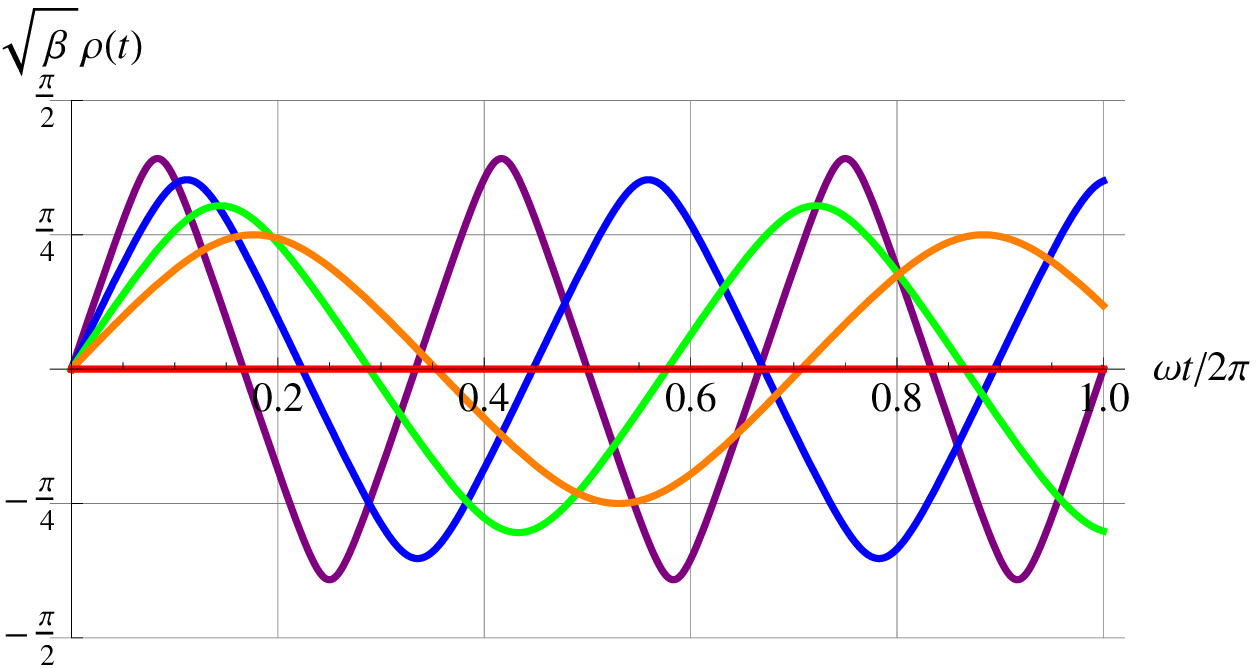}
\includegraphics[width=8cm]{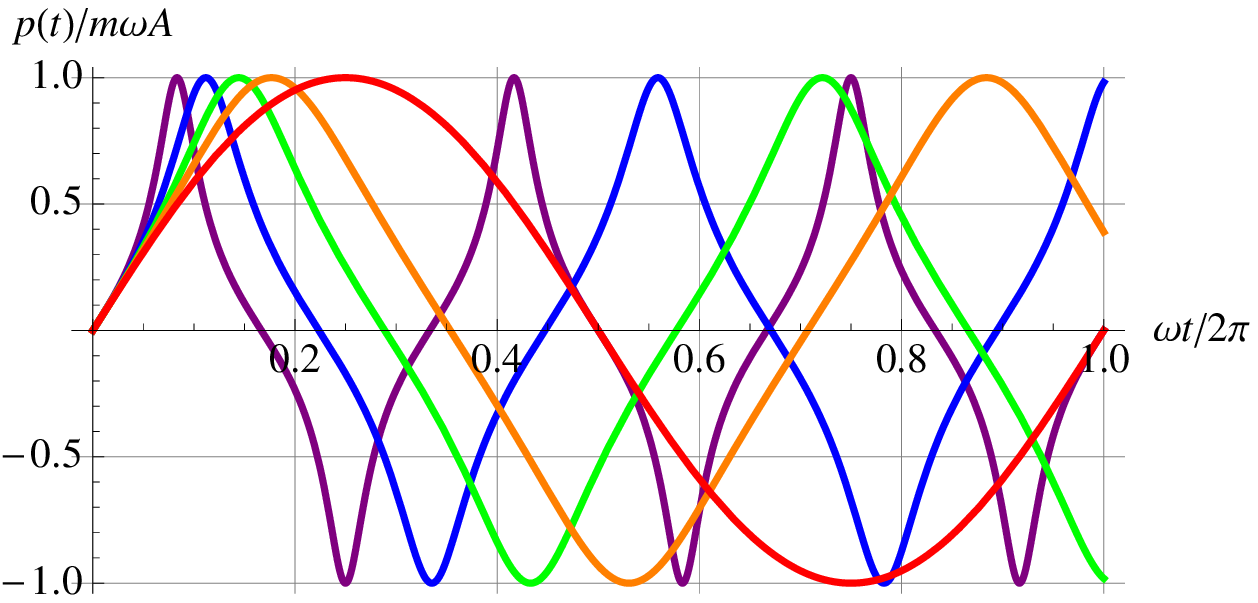}
\includegraphics[width=8cm]{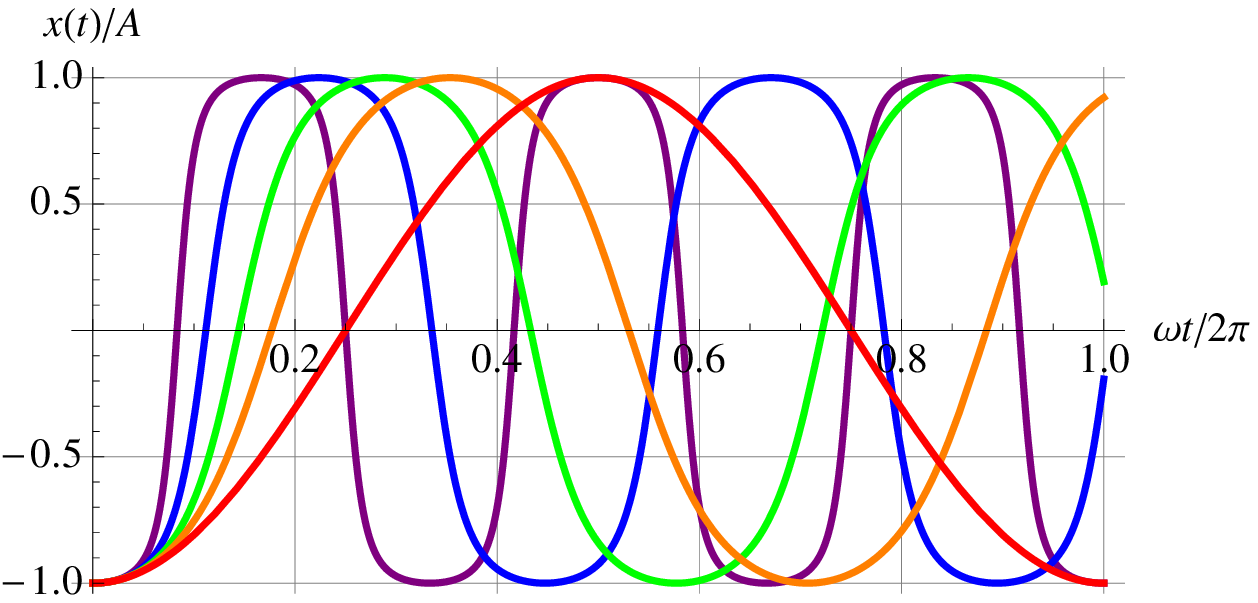}
\end{flushright}
\caption{The dependence of the classical behavior of a positive mass particle in a
harmonic oscillator potential on the parameter $C=2m E\beta=\beta m^2\omega^2 A^2$,
where $E$ is the particle's energy, and $A$ is the amplitude of the oscillation in $x$.
The undeformed $\beta=0$ case is shown in red. The other four cases are $C=1$ (orange),
$C=2$ (green), $C=4$ (blue), and $C=8$ (purple).
}
\label{XplotPplotNorm1}
\end{figure}


\begin{figure}[t]
\begin{flushright}
\includegraphics[width=8cm]{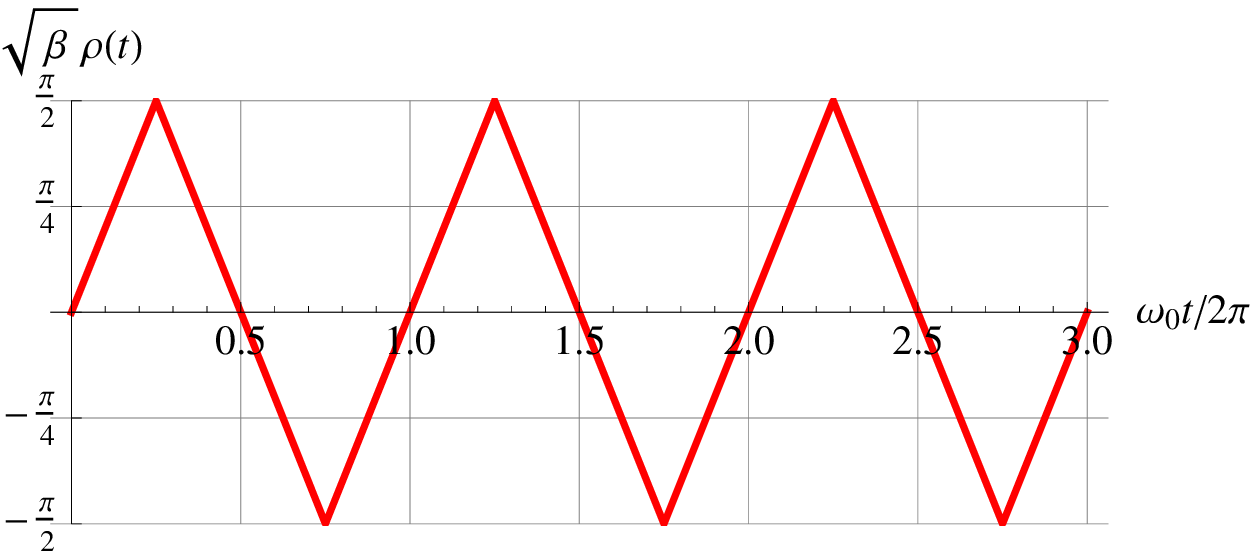}
\includegraphics[width=8cm]{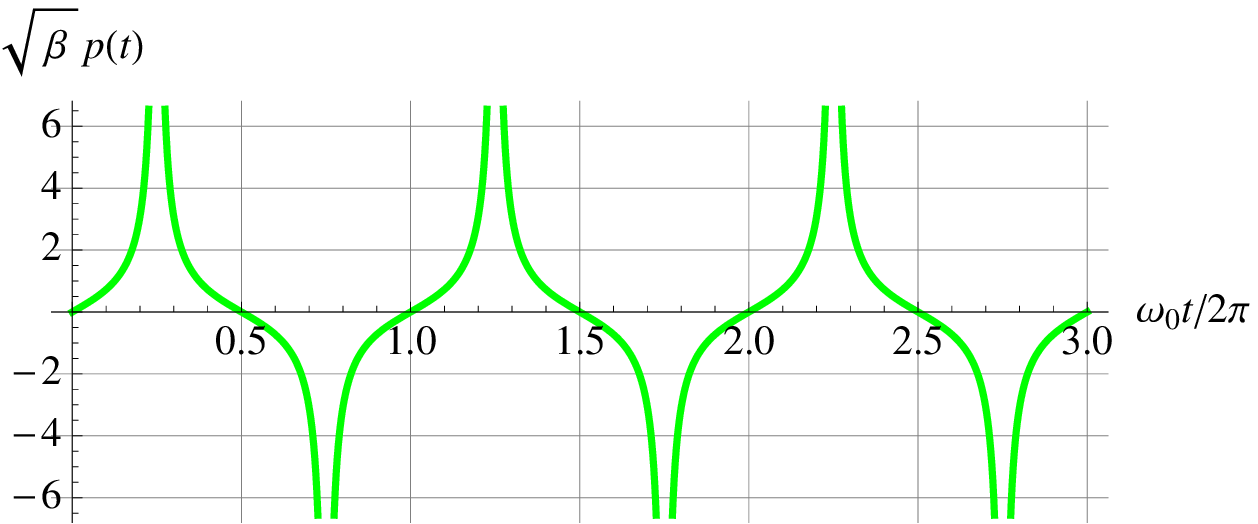}
\includegraphics[width=8.14cm]{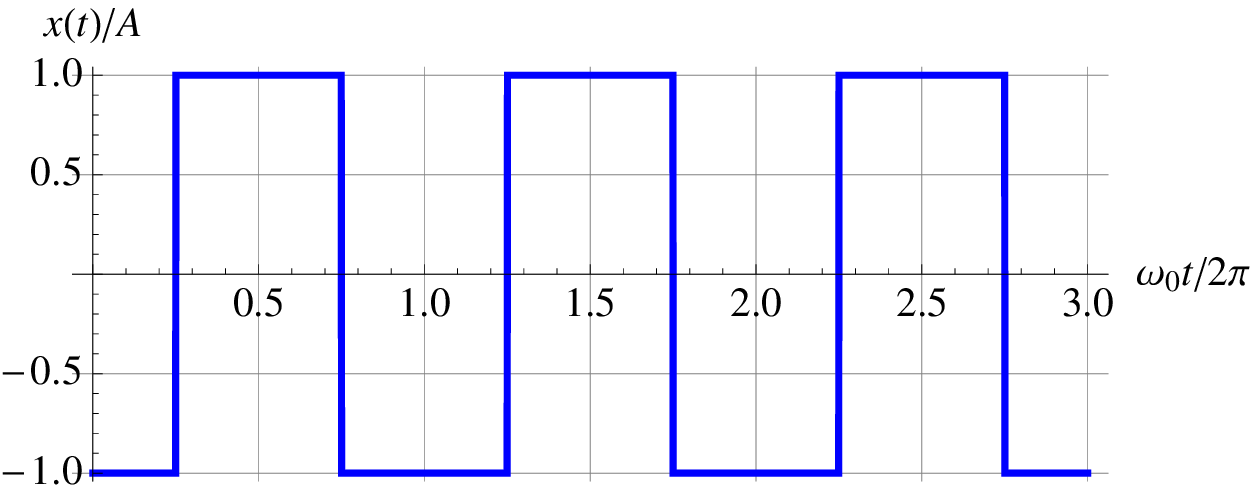}
\end{flushright}
\caption{The classical behavior of a positive mass particle in a harmonic oscillator potential
with modified Poisson brackets,
Eq.~(\ref{ModifiedPoisson}), in the limit $m\rightarrow\infty$ with $C=2mE\beta$
where $E$ and $\beta$ are kept fixed.
$\rho(t)$ and $x(t)$ take on the behavior of the position and momentum of a particle
in an infinite square well.
}
\label{RhoplotXplotPplot}
\end{figure}


\subsection{$\bm{m>0}$ case}

\noindent
When $m>0$, we introduce the angular frequency
\begin{equation}
\omega \;=\; \sqrt{\dfrac{k}{m}}
\end{equation}
as usual. Then,
Eq.~(\ref{rhodotsquared}) becomes
\begin{equation}
\beta\,\dot{\rho}^2 \;=\; 
\omega^2
\biggl[\,C-\tan^2\bigl(\sqrt{\beta}\rho\bigr) \biggr]\;,
\label{rhodotsquared2}
\end{equation}
and taking the square-root, we obtain
\begin{equation}
\sqrt{\beta}\,\dot{\rho} \;=\; \pm\,\omega \sqrt{C - \tan^2(\sqrt{\beta}\rho)}\;.
\end{equation}
In this case, we must have $C> 0$ for the content of the square-root to be
positive.
Separating variables, we obtain
\begin{equation}
\dfrac{\sqrt{\beta}\,d\rho}{\sqrt{C-\tan^2\bigl(\sqrt{\beta}\rho\bigr)}} \;=\; 
\pm\,\omega\,dt\;.
\end{equation}
The left-hand side integrates to
\begin{eqnarray}
\lefteqn{\int\dfrac{\sqrt{\beta}\,d\rho}{\sqrt{C-\tan^2(\sqrt{\beta}\rho)}} 
} \cr
& = &
\dfrac{1}{\sqrt{1+C}}\;\arcsin\!\left[\sqrt{\dfrac{1+C}{C}}\,\sin\bigl(\sqrt{\beta}\rho\bigr)\right]
\;.
\end{eqnarray}
Therefore,
\begin{eqnarray}
\lefteqn{\sqrt{\beta}\rho(t)} \cr
& = & 
\arcsin\!\left[
\sqrt{\dfrac{C}{1+C}}\;
\sin\left\{\pm\sqrt{1+C}\,\omega (t-t_0)\,\right\}
\right]
\cr
& & 
\end{eqnarray}
where $t_0$ is the integration constant.
Without loss of generality, we can choose the sign inside the curly brackets to be plus.
Setting the clock so that $t_0=0$, we obtain:
\begin{eqnarray}
\sqrt{\beta mk}\,x(t) 
& = & -\dfrac{\sqrt{\beta}}{\omega}\,\dot{\rho} \cr
& = & 
-\dfrac{ \sqrt{C(1+C)}\;\cos\bigl(\sqrt{1+C}\,\omega t\bigr) }
        { \sqrt{1+C\,\cos^2\bigl(\sqrt{1+C}\,\omega t\,\bigr)} }
\;,\cr
\sqrt{\beta}\,p(t) 
& = & \tan(\sqrt{\beta}\rho) \cr
& = & 
\dfrac{ \sqrt{C}\;\sin\bigl(\sqrt{1+C}\,\omega t\bigr) }
        { \sqrt{1+C\,\cos^2\bigl(\sqrt{1+C}\,\omega t\,\bigr)} }
\;,
\end{eqnarray}
and the energy is given by
\begin{eqnarray}
E
& = & \dfrac{k}{2}\,x(t)^2 + \dfrac{p(t)^2}{2m}\cr
& = &
 \dfrac{1}{2\beta m}\left[\sqrt{\beta mk}\,x(t)\right]^2 
+\dfrac{1}{2\beta m}\left[\sqrt{\beta}\,p(t)\right]^2
\cr
& = & \dfrac{C}{2\beta m} \;>\; 0\;.
\end{eqnarray}
The period of oscillation $T$ is no longer equal to $2\pi/\omega$ when $\beta\neq 0$.
It is now
\begin{equation}
T\;=\; \dfrac{2\pi}{\omega}\,\dfrac{1}{\sqrt{1+C}}\;.
\end{equation}
Let
\begin{equation}
A\;\equiv\;\sqrt{\dfrac{C}{\beta mk}}\;.
\end{equation}
Then
\begin{equation}
E \;=\; \dfrac{1}{2}kA^2 
\;,
\end{equation}
and we can identify $A$ as the oscillation amplitude in $x$.
If we take the limit $\beta\rightarrow 0$ while keeping $A$ fixed, we find:
\begin{eqnarray}
x(t) & \;\xrightarrow{\beta\rightarrow 0}\; & -A\cos(\omega t)\;,\cr
p(t) & \;\xrightarrow{\beta\rightarrow 0}\; & Am\omega\sin(\omega t)\;,
\end{eqnarray}
which shows that the canonical behavior is recovered in this limit.
The behavior of the solution when $\beta\neq 0$ is compared with the $\beta=0$ limit
for several representative values of $C$ in Fig.~\ref{XplotPplotNorm1}.

Another interesting limit is obtained by setting $C=2mE\beta$
and letting $m\rightarrow\infty$ while keeping $E$ and $\beta$ fixed.
In that limit, 
\begin{equation}
\sqrt{1+C}\,\omega \quad\xrightarrow{m\rightarrow\infty}\quad
\sqrt{2E\beta k}
\;= \sqrt{\beta}kA \;\equiv\;\omega_0\;,
\end{equation}
and we find
\begin{eqnarray}
\sqrt{\beta}\,\rho(t) 
& \;\xrightarrow{m\rightarrow\infty}\; & \arcsin\Bigl[\sin(\omega_0 t)\Bigr]
\;,\cr
\sqrt{\beta}\,p(t)    
& \;\xrightarrow{m\rightarrow\infty}\; & +\dfrac{\sin(\omega_0 t)}{|\cos(\omega_0 t)|}
\;,\cr
x(t)/A 
& \;\xrightarrow{m\rightarrow\infty}\; & -\dfrac{\cos(\omega_0 t)}{|\cos(\omega_0 t)|}
\;.
\label{mPlusInfinityLimit}
\end{eqnarray}
The behavior of the solution in this limit is shown in Fig.~\ref{RhoplotXplotPplot}.
The motion of a particle in an infinite square well potential (in $\rho$-space)
is reproduced, in correspondence to the quantum $\lambda\rightarrow 1$ limit.

\subsection{$\bm{m<0}$ Case}

\noindent
For the $m<0$ case, by an abuse of notation, let us set
\begin{equation}
\iomega \;=\; \sqrt{\dfrac{k}{|m|}}\;.
\end{equation}
Then, Eq.~(\ref{rhodotsquared}) becomes
\begin{equation}
\beta\,\dot{\rho}^2 \;=\; 
\iomega^2\bigg[\tan^2\bigl(\sqrt{\beta}\rho\bigr) - C\,\biggr]\;.
\label{rhodotsquared3}
\end{equation}
The integration constant $C$ can have either sign in this case.
We will consider the three cases $C<0$, $C>0$, and $C=0$ separately.


\begin{figure}[t]
\includegraphics[width=8cm]{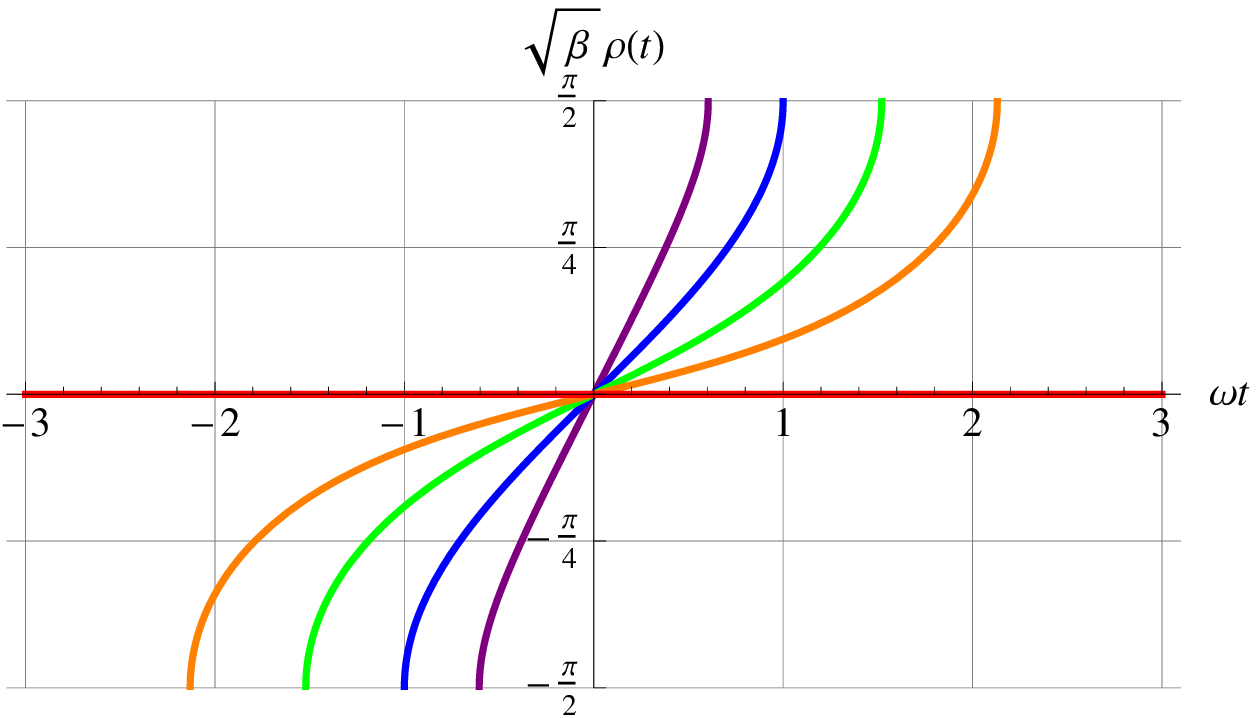}
\includegraphics[width=8cm]{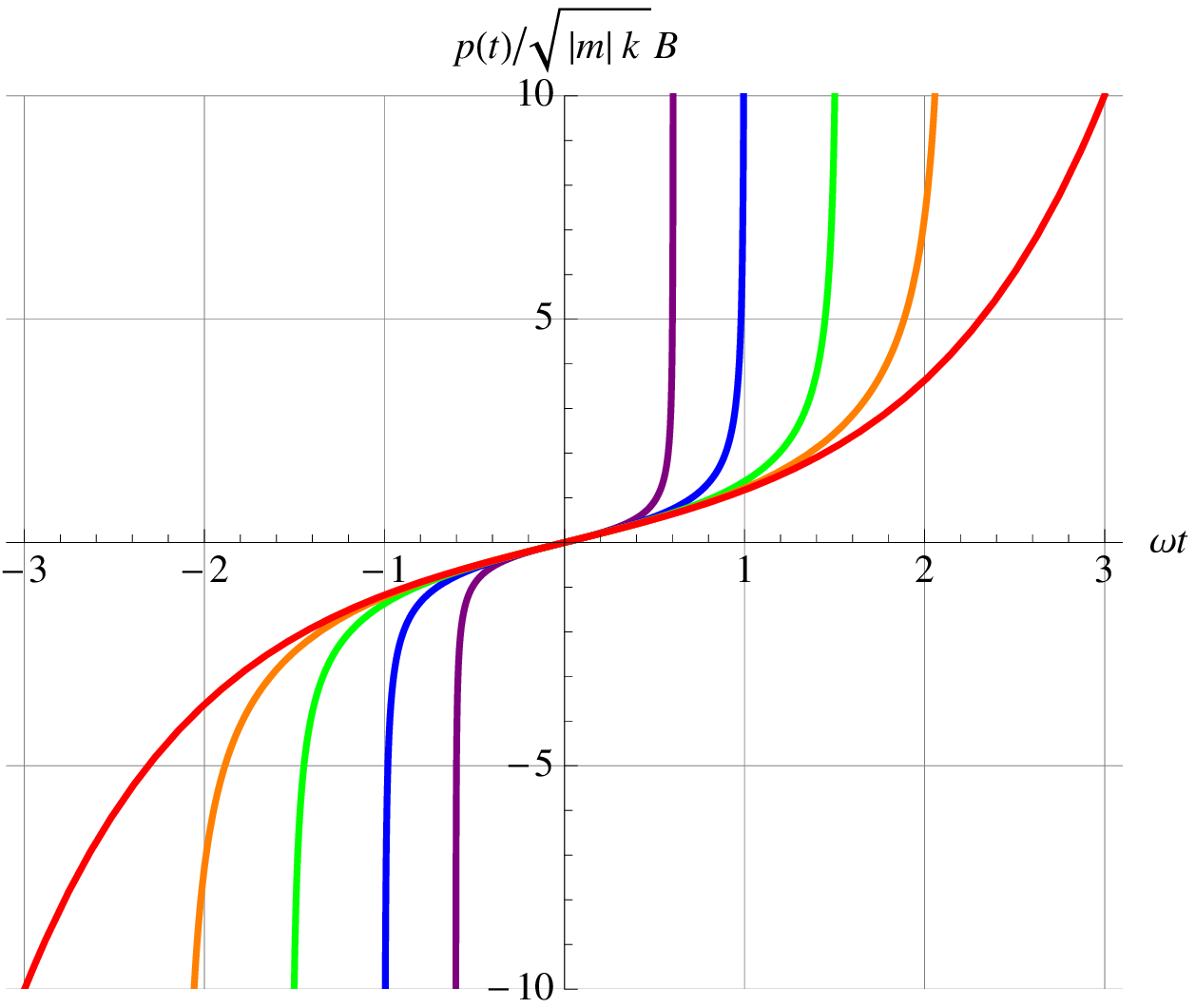}
\includegraphics[width=8cm]{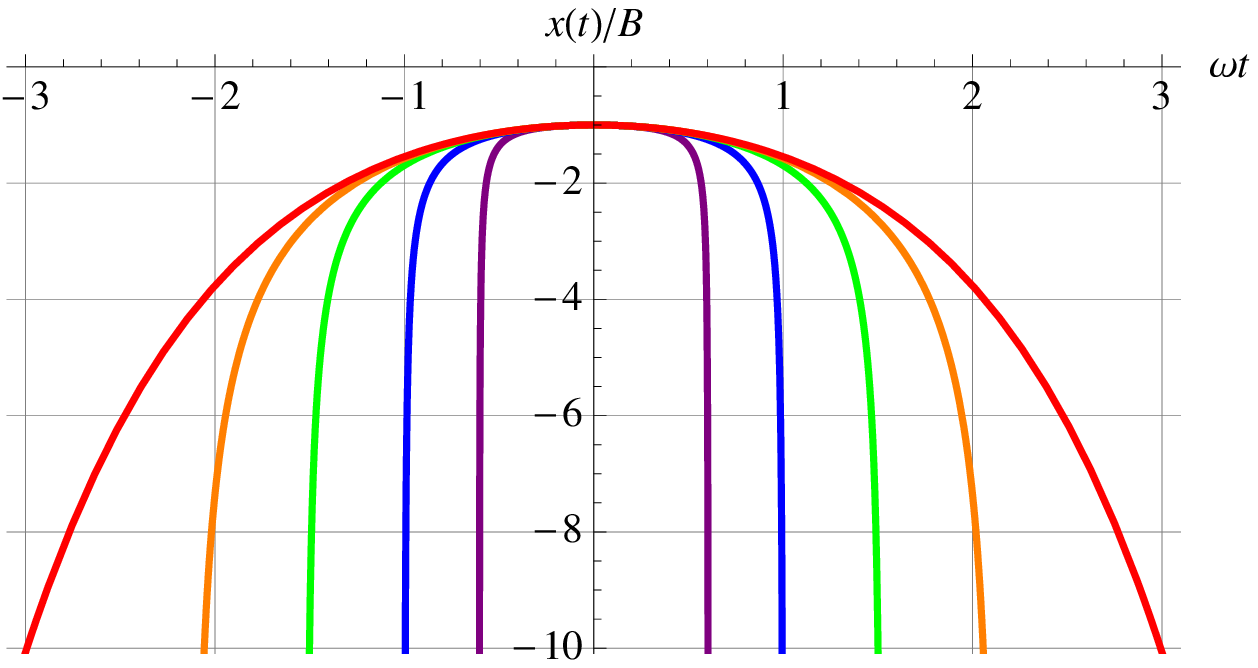}
\caption{The dependence of the classical behavior of a negative mass particle in a
harmonic oscillator potential on the parameter $C=-2\mu E\beta=-\beta|m|k B^2$,
where $E$ is the particle's energy, and $B$ is the distance of closest approach to the origin
in $x$-space.
Note that due to the negative mass, $p(t)$ is negative when $\dot{x}(t)$ is positive, and vice versa.
The undeformed $\beta=0$ case is shown in red. The other four
cases are $C=-\frac{1}{16}$ (orange), $C=-\frac{1}{4}$ (green), $C=-1$ (blue), and $C=-4$ (purple).
}
\label{XplotPplotInvertedCminus}
\end{figure}


\subsubsection{$C < 0$ (positive energy) case}

\noindent
For the $C < 0$ case, the square-root of Eq.~(\ref{rhodotsquared3}) gives us
\begin{equation}
\sqrt{\beta}\,\dot{\rho} \;=\; 
\pm\;\iomega\sqrt{\tan^2(\sqrt{\beta}\rho)+|C|}\;.
\end{equation}
Therefore,
\begin{equation}
\dfrac{\sqrt{\beta}\,d\rho}{\sqrt{\tan^2\bigl(\sqrt{\beta}\rho\bigr)+|C|}} \;=\; 
\pm\;\iomega\,dt\;.
\end{equation}
The left-hand side integrates to
\begin{eqnarray}
\lefteqn{\int
\dfrac{\sqrt{\beta}\,d\rho}{\sqrt{\tan^2(\sqrt{\beta}\rho)+|C|}} 
} \cr
& = & \!\!
\left\{
\begin{array}{ll}
\dfrac{1}{\sqrt{|C|-1}}\,\arcsin\left[\!\sqrt{\dfrac{|C|-1}{|C|}}\sin\bigl(\sqrt{\beta}\rho\bigr)\!\right]
& \bigl(|C|>1\bigr) \\
\;\sin\bigl(\sqrt{\beta}\rho\bigr)\phantom{\bigg|}
& \bigl(|C|=1\bigr) \\
\dfrac{1}{\sqrt{1-|C|}}\,\arcsinh\left[\!\sqrt{\dfrac{1-|C|}{|C|}}\sin\bigl(\sqrt{\beta}\rho\bigr)\!\right]
& \bigl(|C|<1\bigr) \\
\end{array}
\right.
\cr
& &
\end{eqnarray}
Therefore,
\begin{eqnarray}
\lefteqn{\sin\bigl(\sqrt{\beta}\rho(t)\bigr)} \cr
& = & \!\!
\left\{
\begin{array}{ll}
\sqrt{\dfrac{|C|}{|C|-1}}\;
\sin\Bigl[\pm\sqrt{|C|-1}\;\iomega (t-t_0)\Bigr]
& \bigl(|C|>1\bigr) \\
\pm\;\iomega(t-t_0)  \phantom{\bigg|}
& \bigl(|C|=1\bigr) \\
\sqrt{\dfrac{|C|}{1-|C|}}\;
\sinh\Bigl[\pm\sqrt{1-|C|}\;\iomega (t-t_0)\Bigr]
& \bigl(|C|<1\bigr) \\
\end{array}
\right.
\cr
& & 
\end{eqnarray}
where $t_0$ is the integration constant, which we will set to zero in the
following.
From this, we find:
\begin{eqnarray}
\lefteqn{\sqrt{\beta|m|k}\;x(t) 
\;=\; -\dfrac{\sqrt{\beta}}{\iomega}\,\dot{\rho}}\cr
& = & 
\left\{
\begin{array}{ll}
\mp\,\dfrac{ \sqrt{|C|(|C|-1)}\,\cos\bigl(\pm\sqrt{|C|-1}\,\iomega\,t\bigr) }
        { \sqrt{|C|\cos^2\bigl(\pm\sqrt{|C|-1}\,\iomega\,t\,\bigr)-1} }
& \bigl(|C|>1\bigr) \\
\mp
\,\dfrac{1}{\sqrt{1-(\iomega\,t)^2}}
& \bigl(|C|=1\bigr) \\
\mp
\,\dfrac{ \sqrt{|C|(1-|C|)}\,\cosh\bigl(\pm\sqrt{1-|C|}\,\iomega\,t\bigr) }
        { \sqrt{1-|C|\cosh^2\bigl(\pm\sqrt{1-|C|}\,\iomega\,t\,\bigr)} }
& \bigl(|C|<1\bigr) \\
\end{array}
\right.
\cr
\lefteqn{\sqrt{\beta}\,p(t) 
\;=\; \tan(\sqrt{\beta}\rho) 
}\cr
& = & 
\left\{
\begin{array}{ll}
\dfrac{ \sqrt{|C|}\,\sin\bigl(\pm\sqrt{|C|-1}\,\iomega\,t\bigr) }
        { \sqrt{|C|\cos^2\bigl(\pm\sqrt{|C|-1}\,\iomega\,t\,\bigr)-1} }
\qquad\quad
& \bigl(|C|>1\bigr) \\
\pm
\,\dfrac{\iomega\,t}{\sqrt{1-(\iomega\,t)^2}}
& \bigl(|C|=1\bigr) \\
\dfrac{ \sqrt{|C|}\,\sinh\bigl(\pm\sqrt{1-|C|}\,\iomega\,t\bigr) }
        { \sqrt{1-|C|\cosh^2\bigl(\pm\sqrt{1-|C|}\,\iomega\,t\,\bigr)} }
& \bigl(|C|<1\bigr) \\
\end{array}
\right.
\cr
& & 
\label{mMinusEplusSolutions}
\end{eqnarray}
In all three cases, we have
\begin{eqnarray}
E 
& = & \dfrac{k}{2}\,x(t)^2 -\dfrac{p(t)^2}{2|m|} \cr
& = & \dfrac{1}{2\beta|m|}\left[\sqrt{\beta|m|k}\;x(t)\right]^2
-\dfrac{1}{2\beta|m|}\left[\sqrt{\beta}\,p(t)\right]^2 \cr
& = & \dfrac{|C|}{2\beta|m|} \;>\; 0\;.
\label{EclassicalCneg}
\end{eqnarray}
Let
\begin{equation}
B \;\equiv\; \sqrt{\dfrac{|C|}{\beta|m|k}}\;.
\end{equation}
Then
\begin{equation}
E \;=\; \dfrac{1}{2}k B^2 
\;,
\end{equation}
and we can identify $B$ as the distance of closest approach to the origin (aka impact parameter).
Taking the limit $\beta\rightarrow 0$ while keeping $B$ fixed, we find:
\begin{eqnarray}
x(t) & \;\xrightarrow{\beta\rightarrow 0}\; & \mp B \cosh(\pm\iomega t) \;,\cr
p(t) & \;\xrightarrow{\beta\rightarrow 0}\; & B|m|\iomega\sinh(\pm\iomega t) \;,
\end{eqnarray}
which recovers the canonical solution. 
This behavior of $x(t)$ and $p(t)$ for the $\beta=0$ case is compared with that in the
$\beta\neq 0$ case for several representative values of $C$ in
Fig.~\ref{XplotPplotInvertedCminus}.

\begin{figure}[b]
\includegraphics[width=8cm]{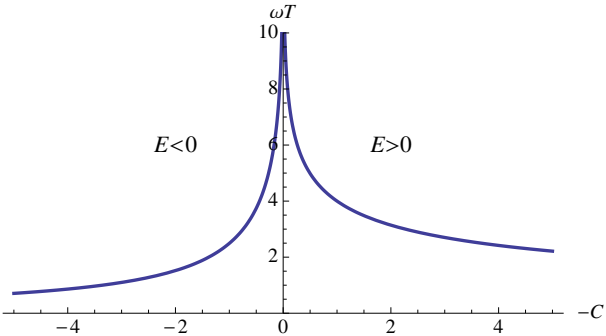}
\caption{$\iomega T$ as a function of $-C=2|m|E\beta$,
where $\iomega=\sqrt{k/|m|}$, and $E$ is the particle's energy.
$T/2$ is the time it takes for the particle to traverse the entire trajectory.
}
\label{OmegaTplotFig}
\end{figure}

It should be noted that for any finite value of $C<0$, it only takes a finite amount of time
for the particle to get from $(x,p)=(\pm\infty,\pm\infty)$ to $(x,p)=(\pm\infty,\mp\infty)$,
or equivalently, for $\sqrt{\beta}\rho$ to evolve from $\mp\pi/2$ to $\pm\pi/2$.  
We will call this time $T/2$ for reasons that will become clear later.
$T$ is given by:
\begin{equation}
T \;=\; \dfrac{4}{\omega}\times
\left\{\begin{array}{cll}
\dfrac{1}{\sqrt{|C|-1}}&\!\!\arccos\dfrac{1}{\sqrt{|C|}} \qquad 
& \bigl(|C|>1\bigr) \\
 1 & \phantom{\bigg|} 
& \bigl(|C|=1\bigr) \\
\dfrac{1}{\sqrt{1-|C|}}&\!\!\arccosh\dfrac{1}{\sqrt{|C|}} 
& \bigl(|C|<1\bigr) \\
\end{array}
\right.
\end{equation}
This dependence on $C<0$ is shown in Fig.~\ref{OmegaTplotFig}.


\begin{figure}[t]
\includegraphics[width=8cm]{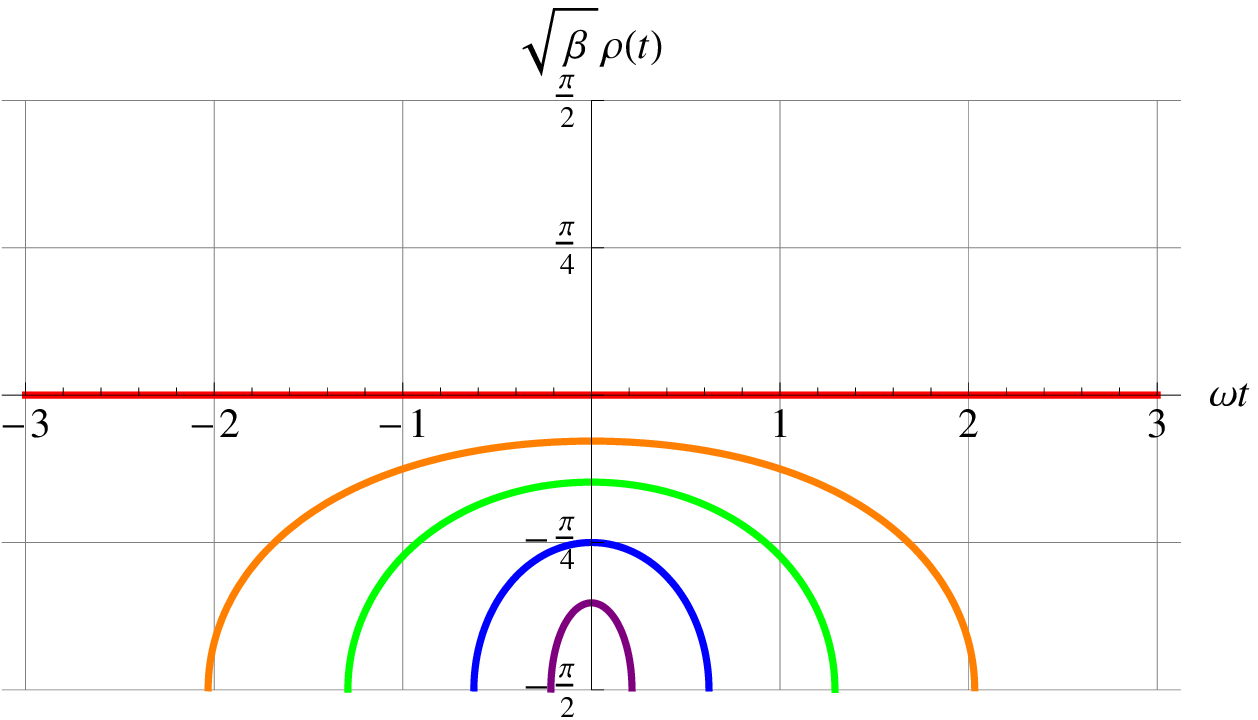}
\includegraphics[width=8cm]{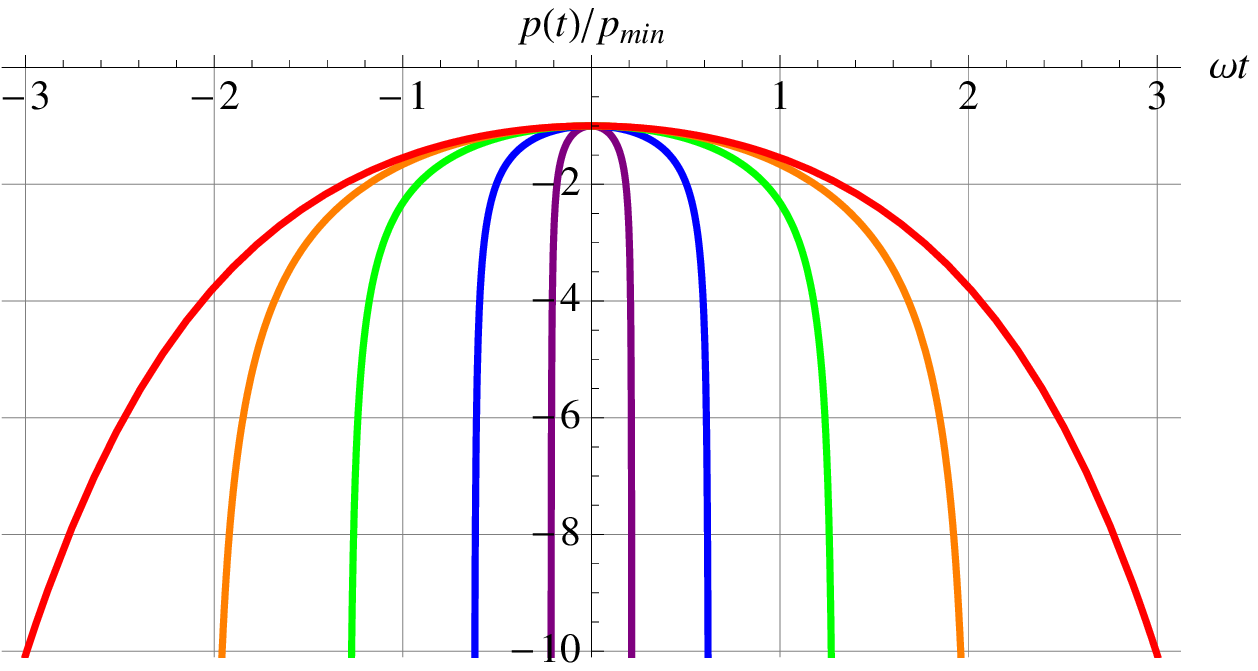}
\includegraphics[width=8cm]{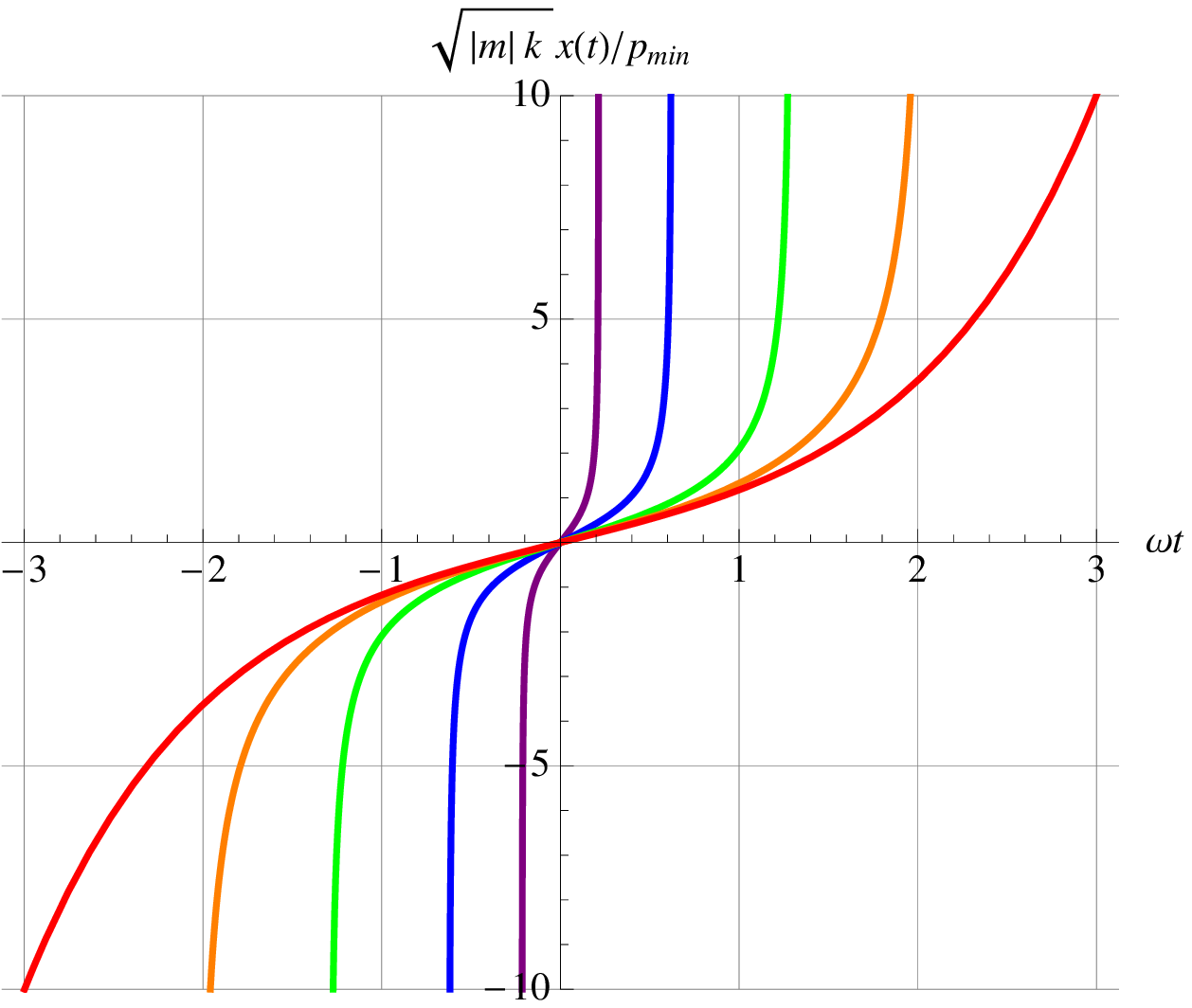}
\caption{The dependence of the classical behavior of a negative mass particle in a
harmonic oscillator potential on the parameter $C=-2|m|E\beta=\beta p_{\min}^2$,
where $E$ is the particle's energy, and $p_{\min}$ is the momentum of the particle at the origin $x=0$.
The undeformed $\beta=0$ case is shown in red. The other four
cases are $C=+\frac{1}{16}$ (orange), $C=+\frac{1}{4}$ (green), $C=+1$ (blue), and $C=+4$ (purple).
}
\label{XplotPplotInvertedCplus}
\end{figure}



\subsubsection{$C>0$ (negative energy) case}

\noindent
For the $C > 0$ case, 
taking the square-root of Eq.~(\ref{rhodotsquared3}) yields
\begin{equation}
\sqrt{\beta}\,\dot{\rho} \;=\; 
\pm\;\iomega
\,\sqrt{\tan^2\bigl(\sqrt{\beta}\rho\bigr)-C}\;.
\end{equation}
Therefore,
\begin{equation}
\dfrac{\sqrt{\beta}\,d\rho}{\sqrt{\tan^2\bigl(\sqrt{\beta}\rho\bigr)-C}} \;=\; 
\pm\,\iomega\,dt\;.
\end{equation}
The left-hand side integrates to
\begin{eqnarray}
\lefteqn{\int
\dfrac{\sqrt{\beta}\,d\rho}{\sqrt{\tan^2(\sqrt{\beta}\rho)-C}} 
} \cr
& = & 
\dfrac{1}{\sqrt{1+C}}\;
\arccosh\!\left|\sqrt{\dfrac{1+C}{C}}\sin\bigl(\sqrt{\beta}\rho\bigr)\right|
\;.
\cr & &
\end{eqnarray}
Therefore,
\begin{eqnarray}
\lefteqn{\sin\bigl(\sqrt{\beta}\rho(t)\bigr)} \cr
& = & 
\pm\sqrt{\dfrac{C}{1+C}}\;
\cosh\Bigl[\pm\sqrt{1+C}\;\iomega (t-t_0)\Bigr]
\end{eqnarray}
where $t_0$ is the integration constant, which we will set to zero in the
following. The sign on the argument of the hyperbolic cosine is also
irrelevant so we will set it to plus.
From this, we find:
\begin{eqnarray}
\sqrt{\beta|m|k}\,x(t) 
& = & -\dfrac{\sqrt{\beta}}{\iomega}\,\dot{\rho}\cr
& = & 
\mp\,\dfrac{ \sqrt{C(1+C)}\,\sinh\bigl(\sqrt{1+C}\,\iomega\,t\bigr) }
        { \sqrt{1-C\sinh^2\bigl(\sqrt{1+C}\,\iomega\,t\,\bigr)} }
\cr
\sqrt{\beta}\,p(t) 
& = & \tan(\sqrt{\beta}\rho) 
\cr
& = & \pm
\dfrac{ \sqrt{C}\,\cosh\bigl(\sqrt{1+C}\,\iomega\,t\bigr) }
        { \sqrt{1-C\sinh^2\bigl(\sqrt{1+C}\,\iomega\,t\,\bigr)} }
\cr & &
\end{eqnarray}
and 
\begin{eqnarray}
E 
& = & \dfrac{k}{2}\,x(t)^2 -\dfrac{p(t)^2}{2|m|} \cr
& = & \dfrac{1}{2\beta|m|}\left[\sqrt{\beta|m|k}\,x(t)\right]^2
-\dfrac{1}{2\beta|m|}\left[\sqrt{\beta}\,p(t)\right]^2
\cr
& = & -\dfrac{C}{2\beta|m|} 
\;<\; 0\;.
\end{eqnarray}
Let
\begin{equation}
p_{\min} \;\equiv\; \sqrt{\dfrac{C}{\beta}}\;.
\end{equation}
Then
\begin{equation}
E \;=\; -\dfrac{p_{\min}^2}{2|m|}\;,
\end{equation}
and we can identify $p_{\min}$ as the magnitude of the momentum that the particle has at the
origin $x=0$.
Taking the limit $\beta\rightarrow 0$ while keeping $p_{\min}$ constant, we find
\begin{eqnarray}
x(t) & \;\xrightarrow{\beta\rightarrow 0}\; & \mp\;\dfrac{p_{\min}}{|m|\iomega}\,\sinh(\pm\iomega t) \;,\cr
p(t) & \;\xrightarrow{\beta\rightarrow 0}\; & \pm\;p_{\min}\,\cosh(\pm\iomega t) \;.
\end{eqnarray}
As in the $C<0$, the time $T/2$ it takes for the particle to travel from 
$x=\mp\infty$ to $x=\pm\infty$ is finite.  $T$ is given by:
\begin{equation}
T \;=\; \dfrac{4}{\iomega\sqrt{1+C}}\;\arcsinh\dfrac{1}{\sqrt{C}}\;.
\end{equation}
This dependence on $C>0$ is also shown in Fig.~\ref{OmegaTplotFig}.


\begin{figure}[t]
\begin{flushright}
\includegraphics[width=8cm]{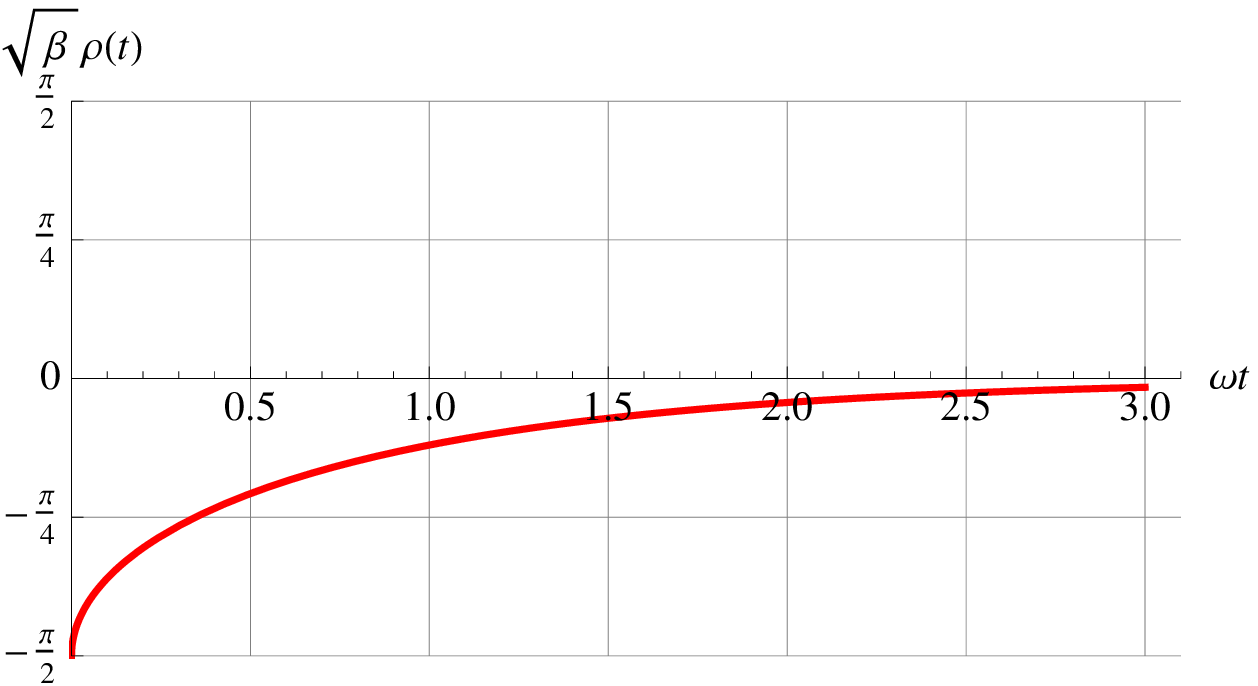}
\includegraphics[width=8cm]{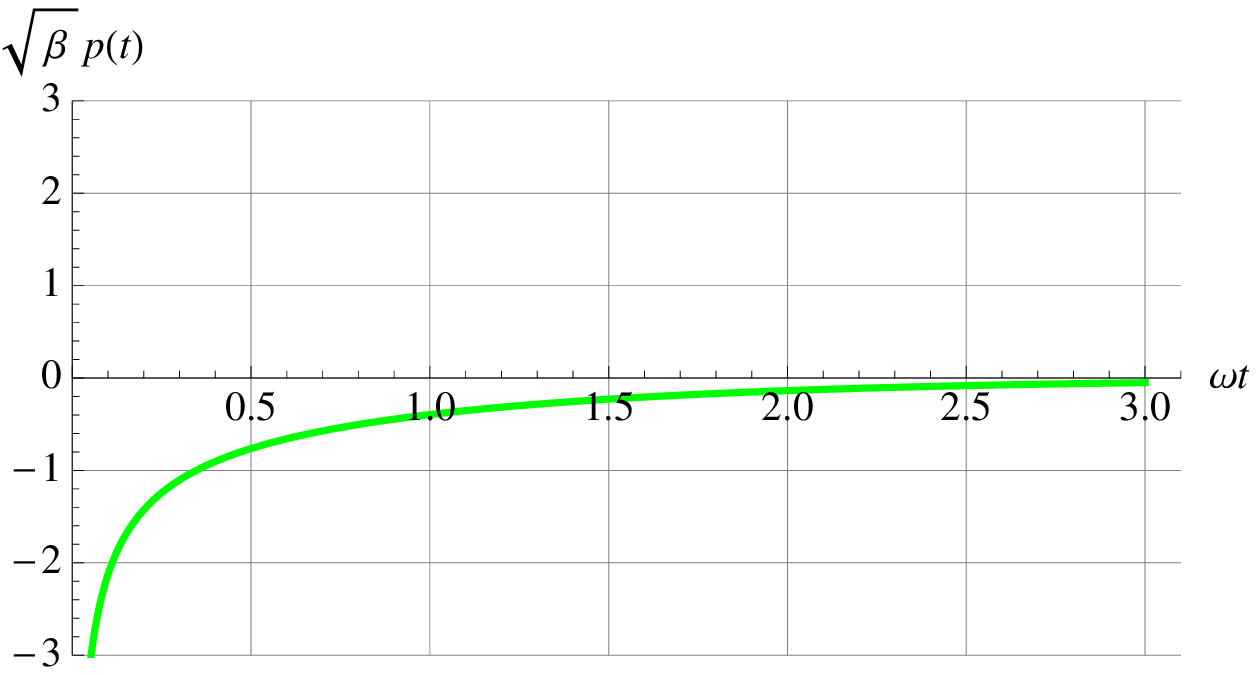}
\includegraphics[width=8.26cm]{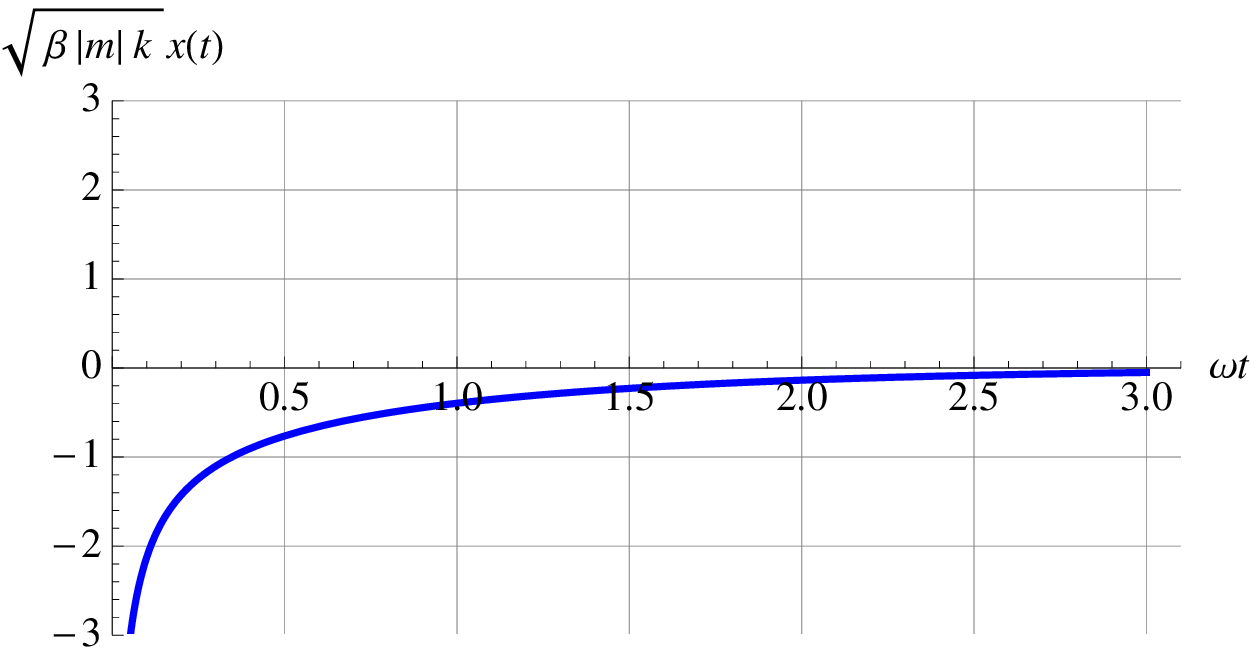}
\end{flushright}
\caption{The classical behavior of a zero-energy particle with negative mass in a
harmonic oscillator potential.
The particle starts out at $x=-\infty$ at time $t=0$ and asymptotically
approaches the origin.
}
\label{XplotPplotInvertedCzero}
\end{figure}


\subsubsection{$C = 0$ (zero energy) case}

\noindent
For $C=0$, Eq.~(\ref{rhodotsquared3}) leads to
\begin{equation}
\sqrt{\beta}\,\dot{\rho} \;=\; \pm\;\iomega\,\tan\bigl(\sqrt{\beta}\rho\bigr)\;,
\end{equation}
or
\begin{equation}
\dfrac{\sqrt{\beta}\,d\rho}{\tan\big(\sqrt{\beta}\rho\bigr)}\;=\; \pm\;\iomega\,dt\;,
\end{equation}
which can be integrated easily to yield
\begin{equation}
\ln\Bigl|\sin\bigl(\sqrt{\beta}\rho\bigr)\Bigr| \;=\; \pm\;\iomega\,(t-t_0)\;,
\end{equation}
or
\begin{equation}
\sin\bigl(\sqrt{\beta}\rho\bigr)
\;=\; \pm\;e^{\,\pm\,\iomega\,(t-t_0)} \phantom{\bigg|}
\;,
\end{equation}
with all combinations of signs allowed.
Set the clock so that $t_0=0$.
The solutions for the $t>0$ region are
\begin{eqnarray}
\begin{array}{rll}
\sqrt{\beta|m|k}\;x(t) 
& = \; -\dfrac{\sqrt{\beta}}{\iomega}\,\dot{\rho} 
& = \; \pm\dfrac{1}{\sqrt{e^{2\iomega t}-1}} \;,\\
\sqrt{\beta}\,p(t)
& = \; \tan\bigl(\sqrt{\beta}\rho\bigr) 
& = \; \pm\dfrac{1}{\sqrt{e^{2\iomega t}-1}} \;.
\end{array}
\end{eqnarray}
The particle starts out at $(x,p)=(\pm\infty,\pm\infty)$ and 
asymptotically approaches the origin, taking an infinite amount of time to get 
there.  This behavior is show in Fig.~\ref{XplotPplotInvertedCzero}.

\bigskip
\subsection{Compactification}


\begin{figure}[t]
\begin{center}
\includegraphics[width=8cm]{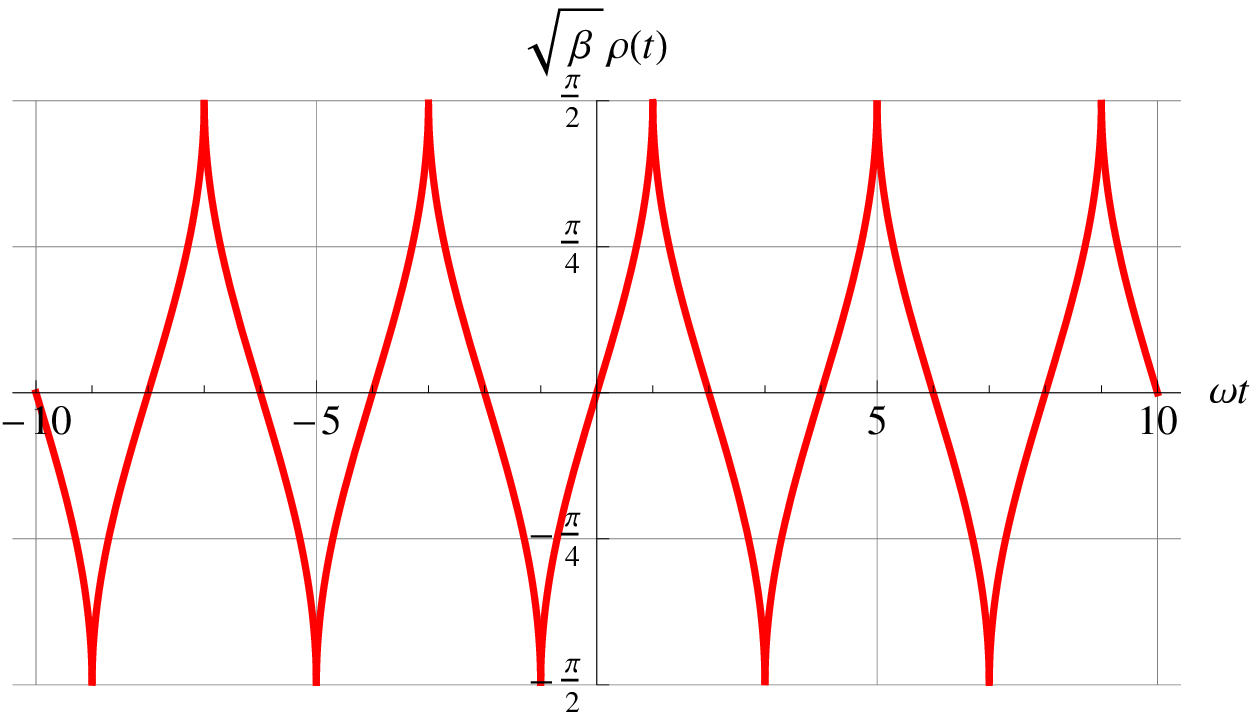}
\includegraphics[width=8cm]{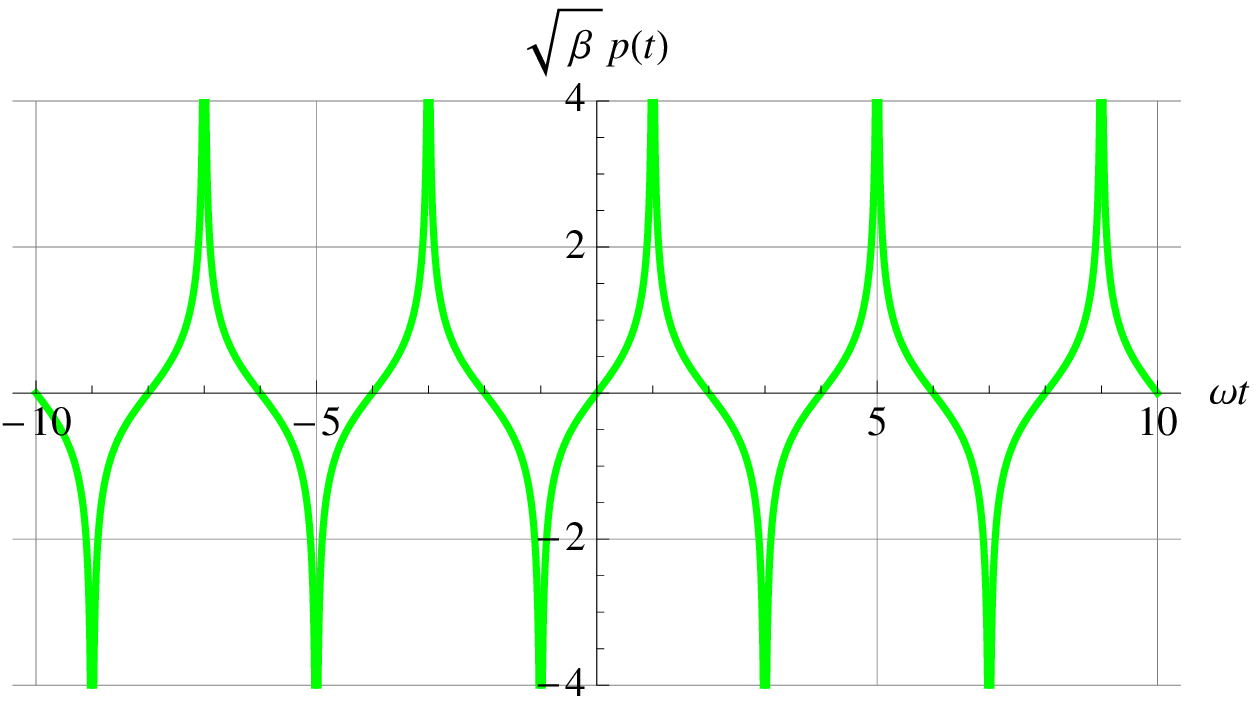}
\includegraphics[width=8cm]{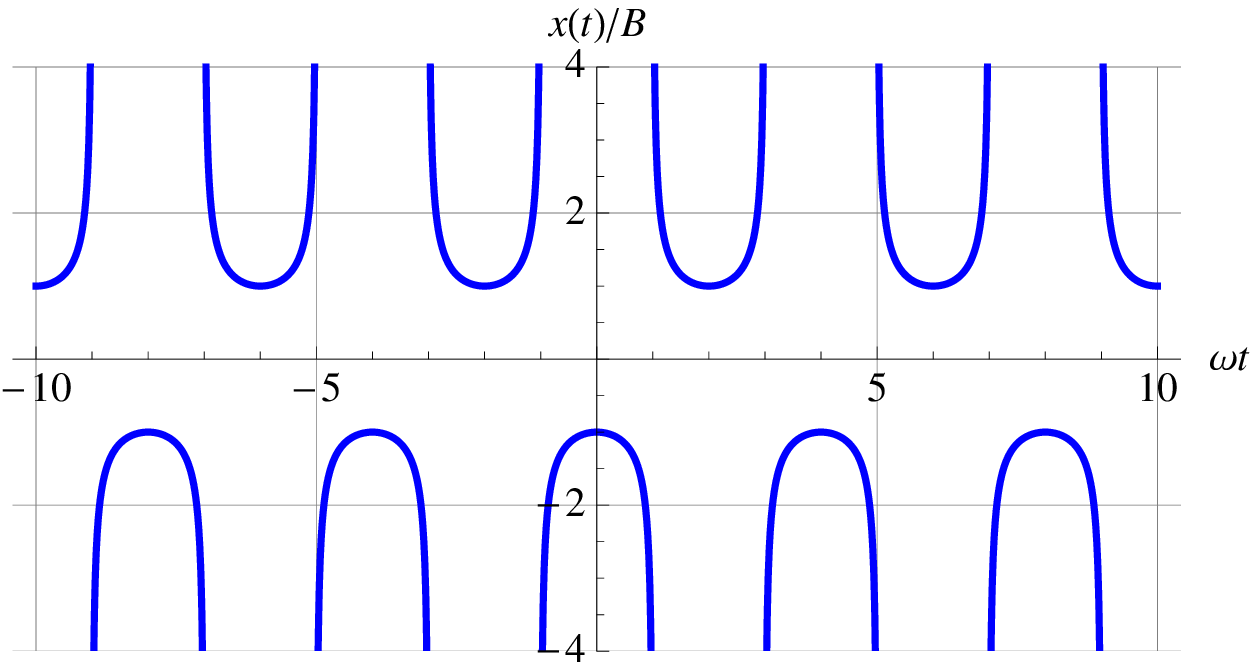}
\end{center}
\caption{The periodic behavior of the negative-mass particle in a
harmonic oscillator potential in compactified $x$-space.
The example shown is for $C=-1$.
}
\label{CompactificationPlot}
\end{figure}


As we have seen, when $m<0$ and $\beta\neq 0$, it only takes a finite amount of time for the
particle to traverse the entire classical trajectory as long as the energy of the particle
is non-zero. 
This means that we must specify what happens to the particle after it reaches infinity.
For this, we could either compactify $x$-space so that the particle
which reaches $x=\pm\infty$ will return from $x=\mp\infty$, in which case the momentum of the
particle will bounce back from infinite walls at $p=\pm\infty$, or we could compactify 
$p$-space so that the particle
which reaches $p=\pm\infty$ will return from $p=\mp\infty$, in which case the position of the
particle will bounce back from infinite walls at $x=\pm\infty$.

Here, we choose to compactify $x$-space so that the $|m|\rightarrow\infty$ limit of the
$m<0$, $C<0$ solution will match the $m\rightarrow\infty$ limit of the 
$m>0$, $C>0$ solution.  
This choice also agrees with the boundary condition we imposed in the quantum case,
in which the wave-function in $\rho$ was demanded to vanish at the domain boundaries
$\rho=\pm\pi/2\sqrt{\beta}$, which corresponds to placing infinite potential walls there.
The $m>0$, $C>0$ solution was given by Eq.~(\ref{mPlusInfinityLimit}).
Taking the $|m|\rightarrow\infty$ limit of Eq.~(\ref{mMinusEplusSolutions})
while keeping $B$ fixed, we find
\begin{equation}
\sqrt{|C|-1}\,\iomega \quad\xrightarrow{m\rightarrow\infty}\quad
\sqrt{2E\beta k}
\;= \sqrt{\beta}kB \;\equiv\;\iomega_0\;,
\end{equation}
and  
\begin{eqnarray}
\sqrt{\beta}\,\rho(t)
& \quad\xrightarrow{|m|\rightarrow\infty}\quad &
\arcsin\Bigl[\sin(\pm\iomega_0 t)\Bigr]
\;,
\cr
\sqrt{\beta}\,p(t) 
& \quad\xrightarrow{|m|\rightarrow\infty}\quad &
+\dfrac{\sin(\pm\iomega_0 t)}{|\cos(\pm\iomega_0 t)|} 
\;,
\cr
x(t)/B 
& \quad\xrightarrow{|m|\rightarrow\infty}\quad &
\mp\dfrac{\cos(\pm\iomega_0 t)}{|\cos(\pm\iomega_0 t)|} 
\;.
\label{mMinusInfinityLimit}
\end{eqnarray}
which formally agrees with Eq.~(\ref{mPlusInfinityLimit}),
and if graphed will lead to a figure similar to Fig.~\ref{RhoplotXplotPplot}.
The one significant difference is, however, that when the particle jumps from 
$x=\pm A$ to $x=\mp A$ in the $m>0$ case it goes through $x=0$, while 
when it jumps from $x=\pm B$ to $x=\mp B$ in the $m<0$ case, it must go through $x=\infty$.

By compactifying $x$-space, all motion when $E\neq 0$ will become oscillatory through $x=\infty$,
and the $T$ calculated above becomes the oscillatory period.
As an example, we plot the $x$-compactified solution for $C=-1$ in Fig.~\ref{CompactificationPlot}, 
for which the period is $T=4/\omega$.
Note that the period for the $m<0$, $C<0$ solution in the limit of $|m|\rightarrow\infty$
becomes
\begin{equation}
\lim_{|m|\rightarrow\infty}T
\;=\; \lim_{|m|\rightarrow\infty}
\dfrac{4}{\omega\sqrt{|C|^2-1}}\arccos\dfrac{1}{\sqrt{|C|}}
\;=\; \dfrac{2\pi}{\iomega_0}\;,
\end{equation}
the arccosine providing a $\pi/2$.

\bigskip
\subsection{Classical Probablities}


\begin{figure}[t]
\begin{flushright}
\includegraphics[width=8cm]{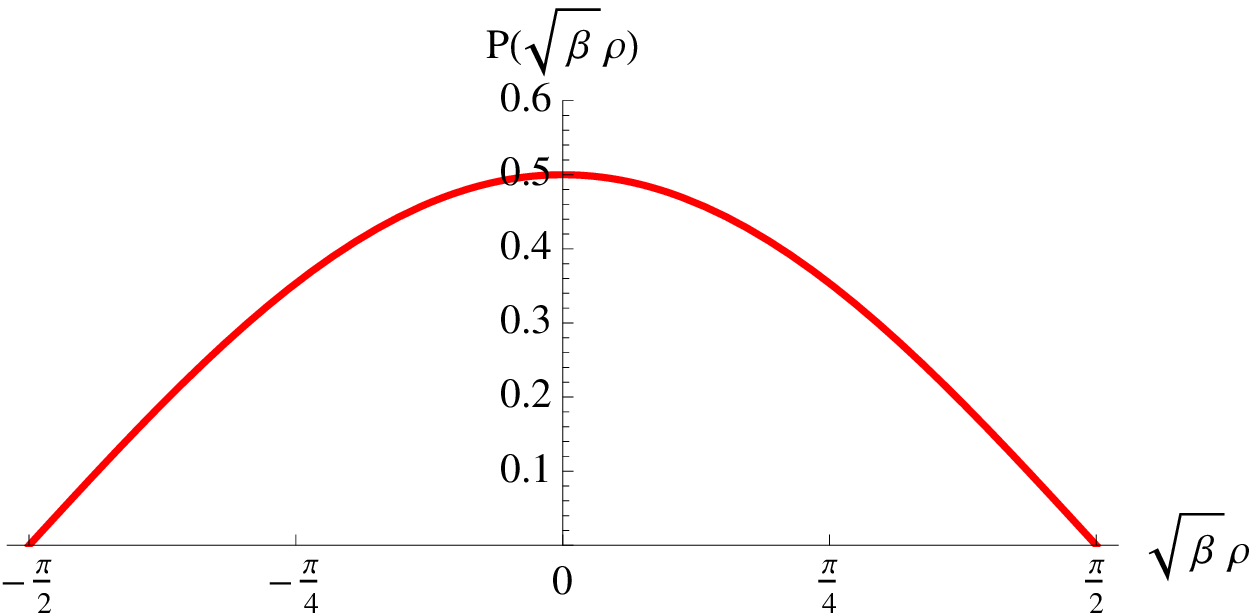}
\includegraphics[width=8cm]{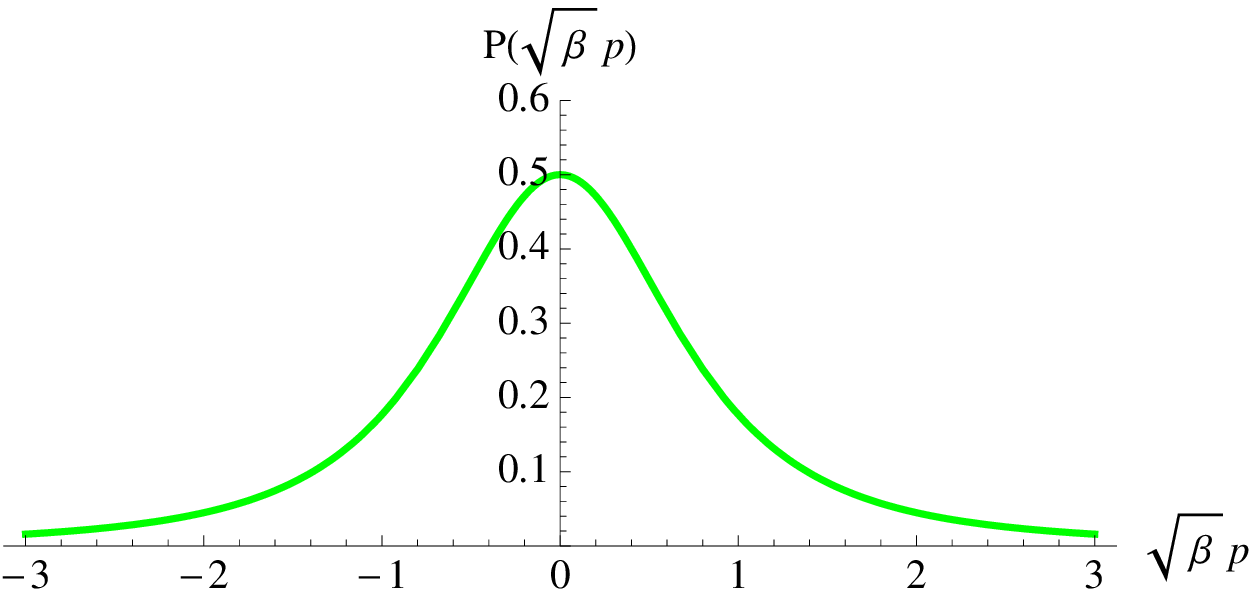}
\includegraphics[width=8.26cm]{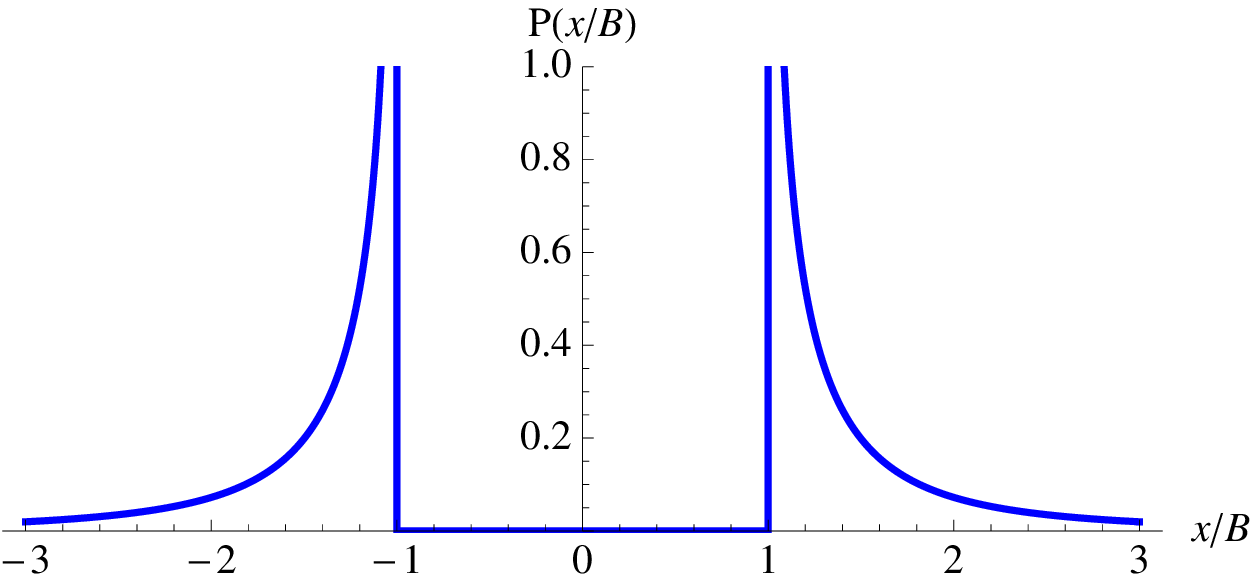}
\end{flushright}
\caption{Classical probabilities in $\rho$-, $p$-, and $x$-spaces 
of a negative mass particle in a harmonic oscillator potential for the case 
$C = -2|m|E\beta = -1$.  
The time dependence of this solution was shown in Fig.~\ref{CompactificationPlot}.
Though the trajectory of the particle is not confined to a finite region
of phase space, the classical probabilities of finding the particle near the
phase-space-origin is still finite due to the finiteness of the time it
takes for the particle to traverse the entire trajectory.
}
\label{ClassicalProbabilityPlot}
\end{figure}


Consider the $m<0$, $C=-2|m|E\beta<0$ case.  
$\sqrt{\beta}\rho$ evolves from $-\pi/2$ to $\pi/2$ in time $T/2$, that is:
\begin{eqnarray}
\dfrac{T}{2} 
& = & \int_{-T/4}^{T/4} dt
\;=\; \int_{-\pi/2\sqrt{\beta}}^{\pi/2\sqrt{\beta}} \dfrac{dt}{d\rho}\,d\rho
\;=\; \int_{-\pi/2\sqrt{\beta}}^{\pi/2\sqrt{\beta}} \dfrac{d\rho}{\dot{\rho}}
\cr
& = & \dfrac{1}{\iomega}
\int_{-\pi/2\sqrt{\beta}}^{\pi/2\sqrt{\beta}}
\dfrac{\sqrt{\beta}\,d\rho}{\sqrt{\tan^2\bigl(\sqrt{\beta}\rho\bigr)+|C|}}
\;.
\end{eqnarray}
Thus, we can identify
\begin{equation}
P(\rho) \;=\;
\dfrac{2}{\iomega T}\;
\dfrac{\sqrt{\beta}}{\sqrt{\tan^2\bigl(\sqrt{\beta}\rho\bigr)+|C|}}
\end{equation}
as the classical probability density of the particle in $\rho$-space.
The classical probablity in $p$- and $x$-spaces can be defined in a similar
manner:
\begin{eqnarray}
P(p) & = & P(\rho)\;\dfrac{d\rho}{dp} \cr
& = & \dfrac{2}{\iomega T}\;
\dfrac{\sqrt{\beta}}{(1+\beta p^2)\sqrt{|C|+\beta p^2}}
\;,\cr
P(x) & = & P(\rho)\;\dfrac{d\rho}{dx} \cr
& = & \dfrac{2}{T}
\dfrac{\sqrt{\beta}|m|}{\sqrt{\beta|m|k x^2-|C|}\Bigl(\beta|m|k^2 x^2 - |C|+1\Bigr)}
\cr
& = & \dfrac{2}{\omega T}
\dfrac{B^2}{\sqrt{x^2-B^2}\Bigl[\,|C|(x^2-B^2)+B^2\,\Bigr]}
\;.
\cr 
& &
\end{eqnarray}
These probability functions are plotted in Fig.~\ref{ClassicalProbabilityPlot}
for the case $C=-2|m|E\beta = -1$.

\begin{figure}[t]
\begin{center}
\includegraphics[width=8cm]{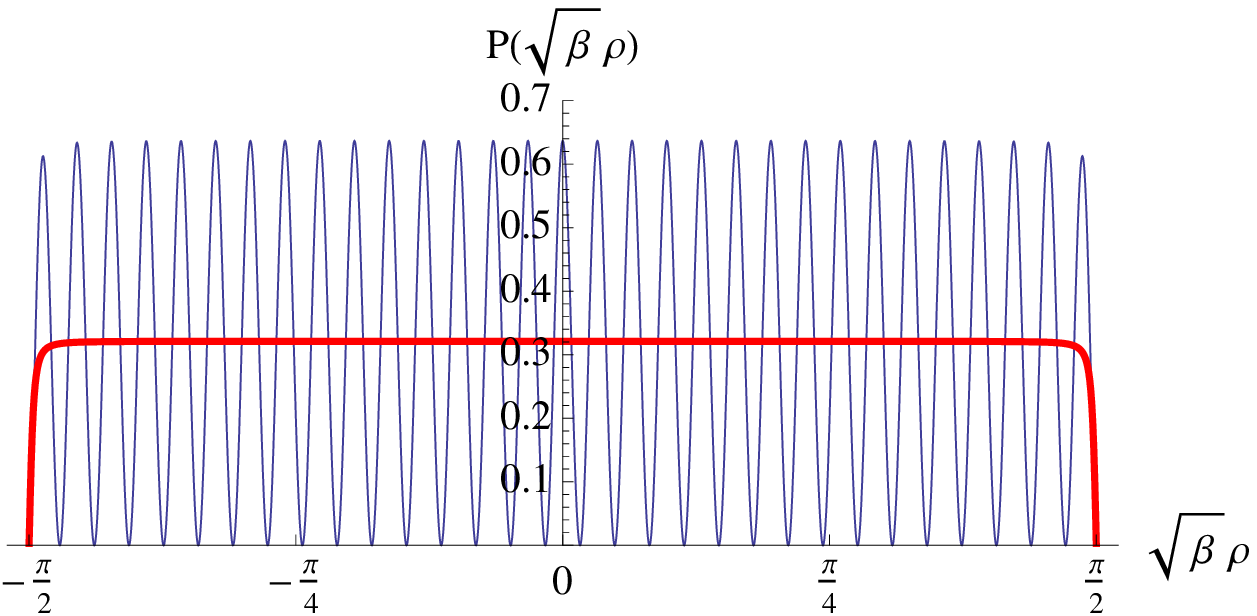}
\includegraphics[width=8cm]{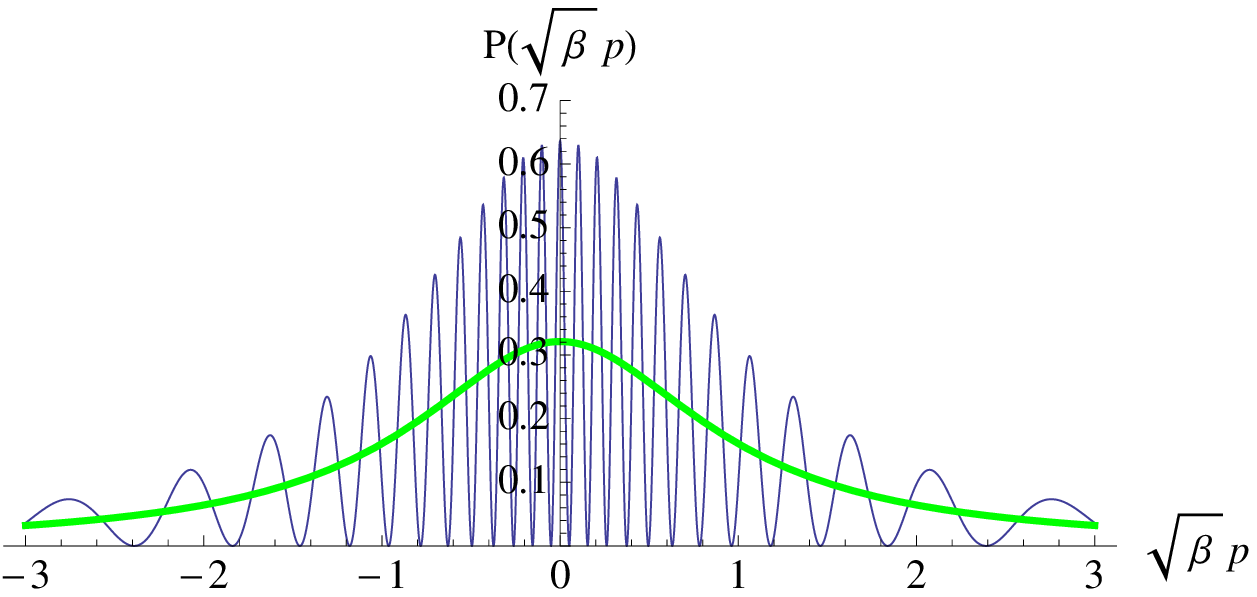}\\
\medskip
\includegraphics[width=8cm]{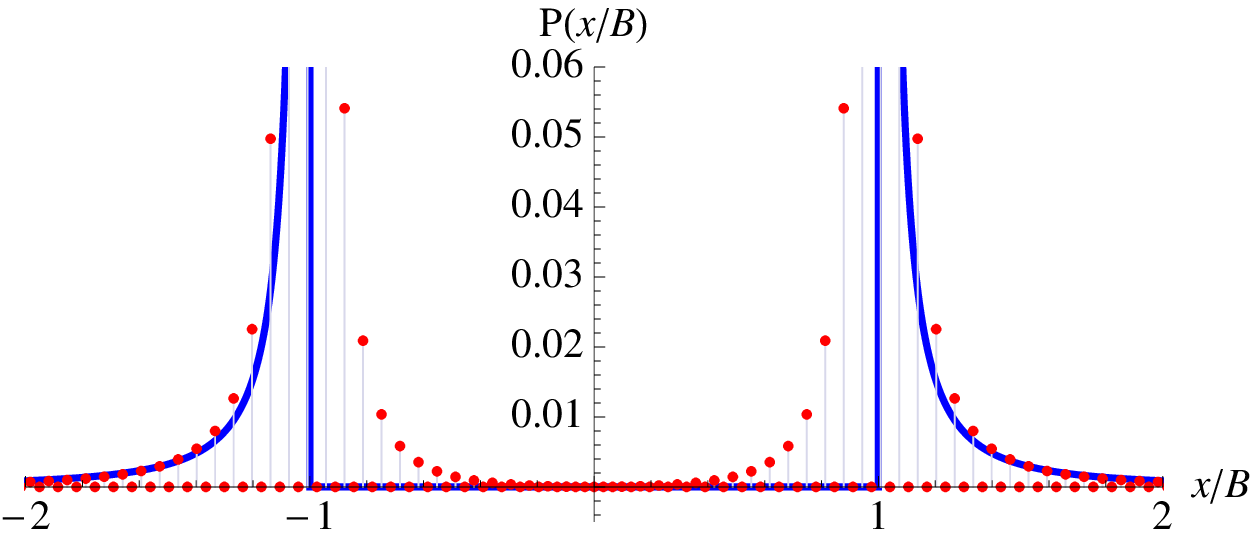}\;\;\;\;\;\\
\end{center}
\caption{Comparison of quantum and classical probabilities in $\rho$-, $p$-, and $x$-spaces 
for a negative mass particle in a harmonic oscillator potential for the case 
$\lambda=\frac{3}{4}$ and $n=30$, which corresponds to the classical $C = -2|m|E\beta = -5044$.  
The quantum distribution in $x$-space is discrete due to the existence of the minimal length.
There is also some seepage of the probability into classically forbidden regions in $x$-space
as is expected of quantum probabilities.
}
\label{ProbabilityComparison}
\end{figure}

Comparing the energies of the quantum and classical solutions given in
Eqs.~(\ref{EnPM}) and (\ref{EclassicalCneg}),
we can conclude that
the correspondence is given by the relation
\begin{equation}
-C \;=\; \dfrac{n^2+(2n+1)\lambda}{\lambda(1-\lambda)}\;,\qquad
\dfrac{1}{2}<\lambda<1\;.
\end{equation}
We expect the quantum and classical probabilities to match for
large $n$.  As an example, we take $\lambda=\frac{3}{4}$ and $n=30$,
which correspond to:
\begin{eqnarray}
C & = & -5044\;,\phantom{\bigg|} \cr
\kappa & = & \dfrac{\Delta x_{\min}}{a} 
\;=\; \dfrac{1}{\sqrt[4]{\lambda(1-\lambda)}} \;=\; \dfrac{2}{\sqrt[4]{3}} \;\approx\;1.52,\cr
B & = & \dfrac{\sqrt{|C|}}{\kappa^2}\,\Delta x_{\min} 
\;\stackrel{1\ll n}{\approx}\; n\,\Delta x_{\min} \;=\; 30\,\Delta x_{\min}\;.
\cr & & 
\end{eqnarray}
The comparison of the quantum and classical probabilities for this case
in $\rho$-, $p$-, and $x$-spaces are shown in Fig.~\ref{ProbabilityComparison}.
If we average out the bumps in the quantum case, it is clear that the
distributions agree, up to the typical quantum mechanical phenomenon of
seepage of the probability into energetically forbidden regions.
Thus, the existence of `bound' states with a finite 
$\Delta x$ and $\Delta p$ in the quantum case can be associated with the
fact that the particle spends a finite amount of time near the phase space origin
in the classical limit.


\section{Summary and Discussion}

We have solved for the eigenstates of the harmonic oscillator hamiltonian
under the assumption of the deformed commutation relation between $\hat{x}$ and $\hat{p}$
as given in Eq.~(\ref{CommutationRelation}), with the objective of calculating their
uncertainties in position and momentum.

For the normal harmonic oscillator with positive mass ($1/m>0$),
the eigenstates are found on the $\Delta x\sim 1/\Delta p$ branch of the MLUR,
where decreasing $1/m$ leads to larger $\Delta p$, and thus smaller $\Delta x$.
Somewhat surprisingly, $1/m$ can be decreased through zero into the negative,
thereby `inverting' the harmonic oscillator, while still maintaining an
infinite ladder of positive energy eigenstates 
so long as the condition $\Delta x_{\min}/\sqrt{2} > a \equiv [\,\hbar^2/k|m|\,]^{1/4}$ 
is satisfied.
There, the eigenstates are found on the $\Delta x\sim\Delta p$ branch of the MLUR,
where decreasing $1/m$ away from zero further into the negative leads to
larger $\Delta p$, and thus larger $\Delta x$, 
with both diverging as $a$ approaches the above bound
from below.
The $1/m=0$ line separating the 
$\Delta x\sim 1/\Delta p$ and $\Delta x\sim \Delta p$ regions is given by 
Eq.~(\ref{MaginotLine}).

Taking the classical limit by replacing our deformed commutator
with a deformed Poisson bracket, we solve the corresponding
classical equations of motion and find that the solutions for the
`inverted' harmonic oscillator are such that the particle only takes
a finite amount of time to traverse its entire trajectory.
This leads to a finite classical probability density of finding 
the particle near the phase space origin, and provides and explanation
of why `bound' states with discrete energy levels are possible 
in the quantum case. 

One significant difference between the classical and quantum cases is, however, 
that the classical system has no restriction on the sign of the energy,
whereas the quantum system only allows for positive energy eigenstates.
The latter is guaranteed by the above mentioned condition on $\Delta x_{\min}$ and $a$. 
Indeed, the condition is equivalent to
\begin{equation}
E \;=\; \dfrac{k}{2}(\Delta x)^2 - \dfrac{1}{2|m|}(\Delta p)^2 \;>\;0\;,
\end{equation}
when one assumes $\Delta x \sim (\hbar\beta/2)\Delta p$.

We have found that the energy eigenstates of the harmonic oscillator
only populate the $\Delta x\sim 1/\Delta p$ branch of the MLUR
when the mass is positive, and only the $\Delta x\sim\Delta p$ branch when 
the mass is negative.  A natural question to ask is whether
some superposition of the energy eigenstates could cross over the
$1/m=0$ line to the other side.  
For the free particle, which can be considered the $k=0$ limit of the harmonic oscillator,
the answer is in the affirmative.
In that case, the uncertainties of the energy eigenstates are $\Delta p=0$ and $\Delta x=\infty$.
Superpositions of these states, with uncertainties positioned anywhere along the MLUR bound,
can be easily constructed, as will be discussed in detail in a
subsequent paper \cite{FreeParticle}.
Is a similar `cross-over' possible for the $k\neq 0$ case?

Another interesting question is whether it is possible to construct classically behaving 
`coherent states' in either of these mass sectors,
and whether their uncertainties are contained as they evolve in time.
In particular, when the mass is negative,
the classical equations of motion calls for the particle to move at
arbitrarily large speeds.
What is the corresponding quantum phenomenon?
We are also assuming that Eq.~(\ref{CommutationRelation}) embodies
the non-relativistic limit of some relativistic theory with a minimal length.
From that perspective, the infinite speed that the negative-mass particle attains
seems problematic.  Would such states not exist in the relativistic theory, or will they
metamorphose into imaginary mass tachyons?
These, and various other related questions will be addressed in future works.

\begin{acknowledgments}

We would like to thank Lay Nam Chang, George Hagedorn, and Djordje Minic for
helpful discussions. 
This work is supported by the U.S. Department of Energy, 
grant DE-FG05-92ER40709, Task A.

\end{acknowledgments}
\appendix
\section{An Integral Formula for Gegenbauer Polynomials}

The following integral formula is necessary in calculating the expectation value of $\hat{p}^2$.
For two non-negative integers $m$ and $n$ such that $m\le n$, and $\lambda>\frac{1}{2}$, we have
\begin{eqnarray}
\lefteqn{
\int_{-1}^{1} 
c^{2\lambda-3}\,C_m^\lambda(s)\,C_n^\lambda(s)\,ds
} \cr
& = & 
\left\{
\begin{array}{ll}
\dfrac{ 4\pi\,\Gamma\bigl(m+2\lambda\bigr) }{ (2\lambda-1)\bigl[\,2^\lambda\Gamma\bigl(\lambda\bigr)\,\bigr]^2 m!}\;
\qquad
& \mbox{if $n-m=\mathrm{even}$,} \\
\phantom{XXX} & \\
0 
& \mbox{if $n-m=\mathrm{odd}$,}
\end{array}
\right.
\cr & &
\label{Integral}
\end{eqnarray}
where $c=\sqrt{1-s^2}$.
We were unable to find this result in any of the standard tables of integrals \cite{SpecialFunctions}
though \textit{Mathematica} seems to be aware of it.
Here we present a proof.

We start from recursion relations which can be found in
Ref.~\cite{SpecialFunctions}.
In section 8.933 of Gradshteyn and Ryzhik, we have:
\begin{eqnarray}
& & (n+2\lambda)\,C_{n}^{\lambda}(x)
\;=\; 2\lambda\Bigl[\, C_{n}^{\lambda+1}(x) - x\,C_{n-1}^{\lambda+1}(x) \,\Bigr]
\;,
\phantom{\Bigg|} 
\cr
& & n\,C_{n}^{\lambda}(x)
\;=\; 2\lambda\Bigl[\, x\,C_{n-1}^{\lambda+1}(x) - C_{n-2}^{\lambda+1}(x) \,\Bigr]
\;.
\end{eqnarray}
Eliminating $C_{n-1}^{\lambda+1}(x)$ and then shifting $\lambda$ by one unit, we obtain
\begin{equation}
C_{n}^{\lambda}(x) - C_{n-2}^{\lambda}(x)
\;=\; \left(\dfrac{n+\lambda-1}{\lambda-1}\right) C_{n}^{\lambda-1}(x)
\;,
\end{equation}
which is Equation 22.7.23 of Abramowitz and Stegun.
Iterating this relation, we deduce that
\begin{eqnarray}
C_{2k}^{\lambda}(x)
& = & C_0^\lambda(x) + \sum_{i=1}^{k}\left(\dfrac{2i+\lambda-1}{\lambda-1}\right) C_{2i}^{\lambda-1}(x)
\;,\cr
C_{2k+1}^{\lambda}(x)
& = & C_1^\lambda(x) + \sum_{i=1}^{k}\left(\dfrac{2i+\lambda}{\lambda-1}\right) C_{2i+1}^{\lambda-1}(x)
\;.
\end{eqnarray}
Since
\begin{eqnarray}
C_0^\lambda(x) & = & 1 \;=\; C_0^{\lambda-1}(x)\;,\cr
C_1^\lambda(x) & = & 2\lambda\,x \;=\; \left(\dfrac{\lambda}{\lambda-1}\right) C_1^{\lambda-1}(x)\;,
\end{eqnarray}
we can write
\begin{eqnarray}
C_{2k}^{\lambda}(x)
& = & \sum_{i=0}^{k}\left(\dfrac{2i+\lambda-1}{\lambda-1}\right) C_{2i}^{\lambda-1}(x)
\;,\cr
C_{2k+1}^{\lambda}(x)
& = & \sum_{i=0}^{k}\left(\dfrac{2i+\lambda}{\lambda-1}\right) C_{2i+1}^{\lambda-1}(x)
\;.
\end{eqnarray}
Thus, the even $C_n^\lambda$'s can be expressed as a sum of the even $C_n^{\lambda-1}$'s,
and the odd $C_n^\lambda$'s as a sum of the odd $C_n^{\lambda-1}$'s.
Invoking the orthogonality relation, Eq.~(\ref{GegenbauerOrthogonalityRelation}),
%
%
which is valid when $\lambda>-\frac{1}{2}$,
it is clear that 
\begin{equation}
\int_{-1}^{1} 
c^{2\lambda-3}\,C_{2k}^\lambda(s)\,C_{2\ell+1}^\lambda(s)\,ds \;=\; 0
\end{equation}
for $\lambda>\frac{1}{2}$.
So for the integral of Eq.~(\ref{Integral}) to be non-zero,
$m$ and $n$ must be both even, or both odd.
For two non-negative integers $k$ and $\ell$ such that $k\le\ell$, we find
\begin{eqnarray}
\lefteqn{\int_{-1}^{1} c^{2\lambda-3}\, C_{2k}^\lambda(s)\,C_{2\ell}^\lambda(s)\,ds}
\cr
& = & 
\int_{-1}^{1} c^{2\lambda-3}
\Biggl[\,
\sum_{i=0}^{k}\left(\dfrac{2i+\lambda-1}{\lambda-1}\right) C_{2i}^{\lambda-1}(s)
\,\Biggr]
\cr
& & \qquad\quad\;\times
\Biggl[\,
\sum_{j=0}^{\ell}\left(\dfrac{2j+\lambda-1}{\lambda-1}\right) C_{2j}^{\lambda-1}(s)
\,\Biggr]ds
\cr
& = & \sum_{i=0}^{k}\sum_{j=0}^{\ell}
\left(\dfrac{2i+\lambda-1}{\lambda-1}\right)
\left(\dfrac{2j+\lambda-1}{\lambda-1}\right)
\cr
& & \qquad\qquad\;\times
\int_{-1}^{1} c^{2\lambda-3}\,C_{2i}^{\lambda-1}(s)\,C_{2j}^{\lambda-1}(s)\,ds
\cr
& = & \sum_{i=0}^{k}
\left(\dfrac{2i+\lambda-1}{\lambda-1}\right)^2
\dfrac{2\pi\,\Gamma(2i+2\lambda-2)}{(2i+\lambda-1)\bigl[\,2^{\lambda-1}\Gamma(\lambda-1)\,\bigr]^2 (2i)!}
\cr
& = & 
\dfrac{2\pi}{\bigl[\,2^{\lambda-1}\Gamma(\lambda)\,\bigr]^2}
\sum_{i=0}^{k}
\dfrac{(2i+\lambda-1)\,\Gamma(2i+2\lambda-2)}{(2i)!}
\;,
\cr
& & \phantom{XXX} \cr
\lefteqn{\int_{-1}^{1} c^{2\lambda-3}\, C_{2k+1}^\lambda(s)\,C_{2\ell+1}^\lambda(s)\,ds}
\cr
& = & 
\int_{-1}^{1} c^{2\lambda-3}
\Biggl[\,
\sum_{i=0}^{k}\left(\dfrac{2i+\lambda}{\lambda-1}\right) C_{2i+1}^{\lambda-1}(s)
\,\Biggr]
\cr
& & \qquad\quad\;\times
\Biggl[\,
\sum_{j=0}^{\ell}\left(\dfrac{2j+\lambda}{\lambda-1}\right) C_{2j+1}^{\lambda-1}(s)
\,\Biggr]ds
\cr
& = & \sum_{i=0}^{k}\sum_{j=0}^{\ell}
\left(\dfrac{2i+\lambda}{\lambda-1}\right)
\left(\dfrac{2j+\lambda}{\lambda-1}\right)
\cr
& & \qquad\qquad\;\times
\int_{-1}^{1} c^{2\lambda-3}\,C_{2i+1}^{\lambda-1}(s)\,C_{2j+1}^{\lambda-1}(s)\,ds
\cr
& = & \sum_{i=0}^{k}
\left(\dfrac{2i+\lambda}{\lambda-1}\right)^2
\dfrac{2\pi\,\Gamma(2i+2\lambda-1)}{(2i+\lambda)\bigl[\,2^{\lambda-1}\Gamma(\lambda-1)\,\bigr]^2 (2i+1)!}
\cr
& = & 
\dfrac{2\pi}{\bigl[\,2^{\lambda-1}\Gamma(\lambda)\,\bigr]^2}
\sum_{i=0}^{k}
\dfrac{(2i+\lambda)\,\Gamma(2i+2\lambda-1)}{(2i+1)!}
\;.
\end{eqnarray}
The sums in the above expressions are given by
\begin{eqnarray}
& & \sum_{i=0}^{k}
\dfrac{(2i+\lambda-1)\,\Gamma(2i+2\lambda-2)}{(2i)!}
\;=\; \dfrac{\Gamma(2k+2\lambda)}{2(2\lambda-1)(2k)!}
\;,\cr
& & \sum_{i=0}^{k}
\dfrac{(2i+\lambda)\,\Gamma(2i+2\lambda-1)}{(2i+1)!}
\;=\; \dfrac{\Gamma(2k+2\lambda+1)}{2(2\lambda-1)(2k+1)!}
\;.\cr
& &
\end{eqnarray} 
These relations can be proved by induction in $k$.
Putting everything together, we obtain Eq.~(\ref{Integral}).

Using this formula, we find the
matrix elements of the operator $\hat{p}^2$ to be:
\begin{eqnarray}
\lefteqn{
\bra{m}\hat{p}^2\ket{n}
\;=\;
\bra{n}\hat{p}^2\ket{m}
}
\cr
& = & 
\left\{
\begin{array}{lr}
\lefteqn{
\dfrac{1}{\beta}
\left[ -\delta_{mn}
+ \dfrac{2\sqrt{(\lambda+m)(\lambda+n)}}{2\lambda-1}
\sqrt{\dfrac{ n!\,\Gamma(2\lambda+m) }{ m!\,\Gamma(2\lambda+n) }}
\right]
}
& \\
& \mbox{for $n-m=\mathrm{even}$, $m\le n$,} \\
\phantom{XXX} & \\
0 \hspace{3cm} & \mbox{for $n-m=\mathrm{odd}$.}
\end{array}
\right.
\cr
& & 
\end{eqnarray}
In particular, the diagonal elements are given by
\begin{equation}
\bra{n}\hat{p}^2\ket{n} \;=\; \dfrac{1}{\beta}\left(\dfrac{2n+1}{2\lambda-1}\right)\;.
\end{equation}
The expectation value of $\hat{x}^2$ is obtained from
\begin{equation}
\dfrac{k}{2}\vev{\hat{x}^2} \;=\; \vev{\hat{H}} - \dfrac{\vev{\hat{p}^2}}{2m}\;.
\end{equation}

\vspace{3cm}

\end{document}